\documentclass[a4paper,11pt]{article}
\pdfoutput=1 

\usepackage{jcappub} 

\usepackage[T1]{fontenc} 

\usepackage{hyperref}
\usepackage{natbib}
\usepackage{subfiles}
\usepackage{aas_macros}
\usepackage{amsmath}
\usepackage{amssymb}
\usepackage[section]{placeins}
\usepackage{graphicx}
\usepackage{booktabs}
\usepackage[normalem]{ulem}
\usepackage{aas_macros}
\useunder{\uline}{\ul}{}

\newcommand{\unit}[1]{\ensuremath{\, \mathrm{#1}}}
\newcommand{\vbar}{\bar{v}}
\newcommand{\vesc}{v_\text{esc}}

\newcommand{\be}{\begin{equation}}
\newcommand{\ee}{\end{equation}}

\newcommand{\Robj}{R}
\newcommand{\Mobj}{M}
\newcommand{\Kn}{\text{Kn}}

\title{The effectiveness of exoplanets and Brown Dwarfs as sub-GeV Dark Matter detectors}

\author[a,1]{C. Ilie,\note{Corresponding author.}}
\author[b]{C. Levy}
\author[a]{and J. Diks}

\affiliation[a]{Department of Physics and Astronomy, Colgate University,\\13 Oak Dr., Hamilton, NY 13346,  U.S.A.}
\affiliation[b]{Physics Department, Harvard University,\\ Cambridge, MA 02138, U.S.A.}

\emailAdd{cilie@colgate.edu}

\abstract{In this work we demonstrate that Dark Matter (DM) evaporation severely hinders the effectiveness of exoplanets and Brown Dwarfs as sub-GeV DM probes. Moreover, we find useful analytic closed form approximations for DM capture rates for arbitrary astrophysical objects, valid in four distinct regions in the $\sigma-m_X$ parameter space. As expected, in one of those regions the Dark Matter capture saturates to its geometric limit, i.e. the entire flux crossing an object. As a consequence of this region, which for many objects falls within the parameter space not excluded by direct detection experiments, we point out the existence of a DM parameter dependent critical temperature ($T_{crit}$), above which astrophysical objects lose any sensitivity as Dark Matter probes. For instance, Jupiters at the Galactic Center have a $T_{crit}$ ranging from $700$~K (for a $3 M_J$ Jupiter) to $950$~K (for $14 M_J$). This limitation is rarely (if ever) considered in the previous literature of indirect Dark Matter detection based on observable signatures of captured Dark Matter inside celestial bodies.}

\begin{document}
\maketitle
\flushbottom

\section{Introduction}
\label{sec:intro}

One of the most profound problems in modern physics is the that of the nature of Dark Matter (DM). The origin of the term can be traced back to the work of Fritz Zwicky in 1933~\cite{Zwicky:1933} who predicted the existence of a non-luminous component of matter in the Coma Cluster of galaxies to explain the measured velocity dispersion. The idea of DM remained dormant in the literature until the 1970s when Rubin and Ford inferred its existence on galactic scales by measuring the rotational speed of H2 gas surrounding the Andromeda Galaxy~\cite{Rubin:1970}. Since then, a plethora of evidence has emerged supporting the DM hypothesis, including gravitational lensing~\cite{Tyson_1998}, the bullet cluster~\cite{Clowe_2006}, and the Cosmic Microwave Background~\cite{Komatsu:2009,Komatsu:2011,Ade:2015,Aghanim:2018}, which predicts that DM composes $\sim 85\%$ of the mass-budget of the universe.~Despite the clear indicators of DM's existence through its gravitational pull at all scales, the exact nature of DM has remained elusive despite the combined efforts from theorists and experimentalists. Various candidates for the form of DM have emerged since its existence was inferred. These include massive compact halo objects~\cite{Mohapatra:2001} and various forms of particle DM, like Axions~\cite{Weinberg:1977ma} and Weakly Interacting Massive Particles~(WIMPs)~\cite{Servant_2003}, to name a few. Massive compact halo objects (MACHOs) have largely been ruled out as dark matter candidates from experimental data and theoretical arguments\footnote{See Sec.~4.1 of~\cite{Freese:2017dm} for a detailed description on the constraints on the model of massive compact halo objects.}, while Axions, WIMPs, and other particle DM candidates have been heavily constrained by direct detection experiments (e.g. ADMX~\cite{Braine_2020}, XENON1T~\cite{Aprile:2019},  PICO60~\cite{Amole:2019fdf}, etc.) which are designed to measure the effect of DM particles on detectors deep underground.

Despite our lack of understanding DM's true form, a clear picture has emerged for the role of DM in the structure of our universe in the standard cosmological model, $\Lambda$CDM. Overdensities of DM form in the early universe after a period of rapid expansion, known as cosmic inflation, which amplify quantum fluctuations in the early universe~\cite{liddle_lyth_2000}. These dense regions of DM, known as halos, provide the gravitational impetus for baryonic matter to collapse and form the first stars~\cite{Bromm:2003}. Through mergers, these DM halos grow, providing the seeds for galaxy formation~\cite{Bromm:2009}. Simulations of DM in the early universe confirm this picture and also posit the existence of a filament-like structure connecting the DM halos~\cite{Wang_2020}. This hierarchical structure formation predicts that most, if not all, galaxies are formed in DM halos that have a rich sub-structure. 

Dark Matter capture is a mechanism for accumulating particle DM within a compact object~\cite{Gould:1987,Bramante:2017,Ilie:2020Comment,Dasgupta:2019juq}. DM particles incident on an object first become accelerated by the object's gravitational field, then, as they transit the object, may collide with its constituents, losing kinetic energy. Provided the particle loses enough energy, it will become gravitationally bound to the object. From there, a number of processes may occur that leaves a DM footprint on the object, including DM-DM annihilation, and kinetic heat transfer, whereby DM collisions with the object's baryons adds energy to the object~\citep[e.g.][]{Baryakhtar:2017}. Capture, and its observational effects, have long been studied and are of serious interest in the search for DM since many physical situations consist of astrophysical objects within dense DM environments, which intuitively produces higher rates of capture. A non-exhaustive list of previous literature on the observational effects of captured dark matter include the study of neutron stars~\cite{Bramante:2017, Ilie:2020Comment,Baryakhtar:2017,Garani:2020}, Population III stars~\cite{Freese:2008cap,Ilie:2020PopIII,Ilie:2021mcms}, white dwarfs~\cite{Dasgupta:2019juq,Horowitz:2020axx}, and, most relevant to this work, exoplanets~\cite{Leane:2020wob}. 

Soon after the formalism for capture was developed~\citep{Spergel:1985, Gould:1987}, it was realized that the reverse process, evaporation, could be relevant for light DM candidates~\cite{Gould:1987evap}. After sufficient time, captured DM thermalizes in the object and through interactions with the stellar medium has a probability of gaining enough energy to escape the gravitational well of the object. In~\cite{Leane:2020wob}, the authors consider self-annihilating models of DM and use the potential observable temperature of exoplanets to generate sensitivity estimates for exoplanets as DM probes. To do so, the authors cut their sensitivity limits at $\sim 10^{-2}$ GeV, claiming that below this mass evaporation becomes relevant enough to erase sensitivity. They support this claim through a calculation of the ``evaporation mass'' which is defined to identify the mass scale at which evaporation becomes relevant. However, as pointed out in~\cite{Garani:2022}, the traditional way to calculate the evaporation mass neglects an important exponential tail in the calculation, which means the evaporation mass calculated in~\cite{Leane:2020wob} should likely be higher. In this work we explicitly confirm this expectation via numerical calculations. 

One of the main goals of this manuscript is to explicitly compute the evaporation rate for Jupiter-like exoplanets (Jupiters) and show how this affects their potential as DM probes for sub-GeV DM. As a result, in Sec.~\ref{sec:DMSensitivity} we produce sensitivity limits that account directly for evaporation and explicitly show how the evaporation mass approximation ($\sim 10^{-2}$ GeV) used in~\cite{Leane:2020wob} is an underestimate. We find instead, that for the objects considered (Jupiters and Brown Dwarfs placed at the Galactic Center) the evaporation mass for is around $\sim 1$ GeV (see Figs.~\ref{fig:SensitivityLimits} and~\ref{fig:SensitivityLimits_BDs}). As such, exoplanets and Brown Dwarfs are quite poor DM probes in the sub-GeV regime. Additionally, we point out an inherent limitation of any astrophysical probe that uses the potential observation of an enhancement from an expected background surface temperature as a signature of Dark Matter. Namely, the existence of a critical temperature~\footnote{For a definition of the critical temperature ($T_{crit}$) see Eq.~\ref{eq:T_Crit}. Its implications with regards to exoplanets as DM probes are discussed in Sec.~\ref{sec:Summary}}  above which all DM constraining power is lost. This effect was pointed out by us before in~\cite{Ilie:2020Comment}, and it is a result of the saturation of the capture rates in a certain region of the $\sigma$ vs $m_X$ parameter space. Additionally, in Sec.~\ref{sec:Capture} we present  simple analytic approximations  of the DM capture rates that are valid in four distinct regions of the same parameter space (see Eqns.~\ref{eq:CtotRI}-\ref{eq:CtotRIV} and Fig.~\ref{fig:AnalyticCapture_Schematic}). We thus extend our findings of~\cite{Ilie:2020PopIII}, where we considered the case of Pop~III stars, when a definite hierarchy between the escape velocity at the surface of the star and the DM dispersion velocity exists ($\vesc\gg\vbar$).

In this work, we use polytropes to approximate the structure of Jupiters and Brown Dwarfs as this information is necessary to compute the rate of DM evaporation. The polytropic model assumes a pressure-density relation of the form $P = K\rho^{1+\frac{1}{n}}$, where $n$ is the polytropic index. Polytropes have long been used to estimate the structure of self-gravitating gaseous bodies~\cite{Chandrasekhar1939} and is thus a good candidate model for gaseous exoplanets. For Jupiters, we follow~\cite{Garani:2022} and take the index as $n=1$ which is justified by comparing to expected models of Uranus and Neptune~\cite{Helled_2010}, Saturn~\cite{MARLEY2014743} and Jupiter~\cite{Stevenson}. For Brown Dwarfs we take index $n=1.5$, a choice justified in detail by~\cite{Chabrier_2000} or~\cite{Garani:2022}. In Appendix~\ref{sec:analyticApproximations} we include useful approximations of integrals involving DM profiles in the limit of high DM mass ($m_X \gg 1\unit{GeV}$) in objects that are modelled by polytropes. These approximations are very useful for computing the rate of DM annihilation in objects for Supermassive DM since numerical integration procedures become increasingly difficult for the highly cored profiles associated with heavy DM. 

Perhaps one of the most intriguing and counter-intuitive findings of our work is the fact that under certain conditions, dark matter has a ``floating distribution'' towards the surface of the object (see Sec.~\ref{sec:DMDistributions}, and in particular Fig.~\ref{fig:DMProfiles}). This confirms similar results independently found in~\cite{Bramante:2022} and~\cite{Leane:2022}.

\section{Dark Matter processes in exoplanets}
In this section, we briefly review three key Dark Matter processes: capture, evaporation, and annihilation. The interplay between those will controll the effectiveness of any given astrophysical object as a Dark Matter probe. If we label $C$ as the rate at which DM particles accumulate in the object through down-scattering, $\Gamma_A$ the rate at which they annihilate within the object, and $E$ the per-particle rate at which they leave the object through up-scattering, then the differential equation governing the total number of DM particles $N_X$ in the object at time $t$ is:
\begin{equation}
    \frac{dN_X}{dt} = C - \Gamma_A (N_X) - E N_X.
    \label{eq:dNx_DiffEq}
\end{equation}
The annihilation rate, $\Gamma_A (N_X)$, is a model-dependent quantity. In this work, we primarly consider the ``Co-SIMP''~\cite{Smirnov:2020} sub-GeV model, where annihilation is given by: DM$+$DM$+$SM$\rightarrow$ DM$+$SM. In making this choice we follow~\cite{Leane:2020wob}, the original paper that pointed out the usefulness of exoplanets as sub-GeV DM probes. Co-SIMPs are a sub-GeV DM model that evade naturally the Lee-Weinberg~\citep{Lee:1977} lower bound on a thermal relic DM particle mass. We also consider general models that annihilate like DM$+$DM$\rightarrow$ SM$+$SM for high-mass DM when we discuss the approximation of integrals involving highly-cored DM profiles. Both models obey $\Gamma_A \sim N_X^2$ since only two DM particles are required. By defining $C_A$ as an $N_X$-independent annihilation coefficient, we can write the annihilation rate as:
\begin{equation}
    \Gamma_A(N_X) = C_A N_X^2,
\end{equation}
With this in hand, a closed-form solution for Eq.~\ref{eq:dNx_DiffEq} can be found:
\begin{equation}
    N_X(t) = \sqrt{\frac{C}{C_A}} \frac{\tanh\left(\frac{\kappa t}{\tau_{eq}}\right)}{\kappa +\frac{1}{2} E \tau_{eq} \tanh\left(\frac{\kappa t}{\tau_{eq}}\right)},
    \label{eq:Nx_t_Solved}
\end{equation}
where $\tau_{eq} \equiv \left(C C_A\right)^{-1/2}$ and $\kappa \equiv \sqrt{1 + \left(\frac{1}{2} E \tau_{eq}\right)^2}$. One can verify that when $t\gg \frac{\tau_{eq}}{\kappa}$, equilibrium is achieved and:
$$N_X \simeq \sqrt{\frac{C}{C_A}} \frac{1}{\kappa + \frac{1}{2} E \tau_{eq}}.$$ 
In the following subsections, we outline in more details the processes of capture, annihilation, and evaporation and how to compute them in the context of exoplanets.

\subsection{Capture}\label{sec:Capture}
Dark Matter capture is the process through which DM particles incident on an astrophysical object become gravitationally bound as they lose energy transiting this object. First studied in the context of the sun~\cite{Spergel:1985} and the Earth~\cite{Gould:1987} for DM that becomes captured after a single scattering event, and the Earth for DM that requires multiple scattering events~\cite{Gould:1992ApJ}, this process has since gained a lot of attention within the astroparticle physics community for its ability to constrain DM in regions of parameter space inaccessible to Earth-based detection experiments. An important extension of Gould's original DM capture at finite optical depth formalism~\cite{Gould:1992ApJ} came recently with the work of Bramante~\cite{Bramante:2017},\footnote{Also note the comment paper \cite{Ilie:2020Comment} in which several errors in the original~\cite{Bramante:2017} are addressed, and most importantly, the existence of critical temperature that limits the usefulness of astrophysical objects as DM probes whenever $T_{eff}\gtrsim T_{crit}$. Regarding this point see Eq.~\ref{eq:T_Crit} and its discussion, and Sec.~\ref{sec:Summary}} which we use throughout this work, and outline below. This extension provides a convenient way to compute the capture rate in the multiscatter limit.

We also present analytic approximations of the capture rate and discuss the form of the capture rate in various regions of $\sigma-m_X$ parameter space (where $\sigma$ is the DM-proton scattering cross-section), expressions which are derived in Appendix~\ref{sec:AnalyticCapture}. A significant new finding of this work is an extension of the analytic capture rates according to regions presented in~\cite{Ilie:2020PopIII}, where the objects of interest were Pop~III stars. As such, the limit of large escape velocity ($\vesc$) relative to to the DM velocity dispersion~($\vbar$) in the surrounding halo, i.e. the limit of $\vesc \gg \vbar$, was considered. In this work, we relax this assumption and explore the parameter space without any apriori assumption on a hierarchy between $\vesc$ and $\vbar$. Our results when the escape velocity is high agree exactly with the findings of~\cite{Ilie:2020PopIII}. In the limit of $\vbar \gg \vesc$, we find a slightly different picture: one of the regions becomes erased (Region III), and another region is pushed upwards (Region II). We also find that, unlike when $\vesc \gg \vbar$, the boundary of Region I and Region II no longer coincide and a new region between them develops. More details on these new findings can be found in~\cite{Ilie:2023Capture}, though we summarize them below and outline the key steps in Appendix~\ref{sec:AnalyticCapture}.

Capture in its schematic form can be represented by the following differential form:
\begin{equation}
    \frac{dC}{dV d^3u} = \Omega(n_T(r), w(r), \sigma, m, m_X)dF(n_X, u, v_\text{obj}, \vesc^\text{halo}),
\end{equation}
where $dC$ is the differential capture rate per volume $dV$ of the object and velocity $d^3 u$ of the DM particle far away, $\Omega$ is the probability of capture, and $dF$ is the incident flux of DM on the object. Further, $n_T(r)$ is the number density of target nuclei of mass $m$ at radius r,\footnote{From now on, $n_T$ will represent the average number density of the most dominant nuclear species throughout the object. For Jupiters, the most abundant species is hydrogen, making up about $84\%$ of the object~\cite{Guillot:2009ij}. In this work, we assume the object is made entirely of hydrogen for simplicity, though in this case it makes no significant difference. For a full treatement of the capture rate for an arbitrary number of components, see~\cite{Ilie:2021mcms}.} $w(r)$ is the speed of the accelerated DM particles at radius r, $\sigma$ is the DM-nucleon scattering cross section, $m_X$ is the mass of the DM particle, $n_X$ is the number density of DM particles in the surrounding medium, $v_\text{obj}$ is the velocity of the object with respect to the DM halo, and $\vesc^\text{halo}$ is the escape velocity of the DM halo. Computing the capture rate for a spherical object then, in principle, amounts to integrating over the capture rate per shell of the object, integrating over the flux of dark matter particles, and integrating over all DM velocities far from the object. In practice, a number of simplifying assumptions are made to make the problem more digestible, all of which can be shown as valid in the context of exoplanets, as outlined in~\cite{Leane:2020wob}.

Following~\cite{Bramante:2017,Ilie:2020Comment}, we compute the capture rate, $C$ in the following way:
\begin{equation}
    C = \sum_{N = 1}^\infty C_N,
    \label{eq:CtotBasic}
\end{equation}
where $C_N$ is the so-called ``partial capture rate,'' i.e. the capture rate for exactly $N$ scattering events. With the assumptions of~\cite{Bramante:2017}, the partial capture rate is given by:
\begin{equation}
   C_N = \underbrace{\pi \Robj^2}_\textrm{capture area}\times \,\underbrace{n_X \int_0^{\infty} \dfrac{f(u)du}{u}\,(u^2+\vesc^2)}_\textrm{DM flux}\times \, \underbrace{p_{ N}(\tau)}_\textrm{prob. for $N$ collisions}\times \, \underbrace{g_{ N}(u)}_\textrm{prob. of capture},
\end{equation}
where $\Robj$ is the radius of the object, $u$ is the speed of the DM particle far from the object, $f(u)$ is the DM speed distribution in the medium, $\vesc \equiv \sqrt{\frac{2 G \Mobj}{\Robj}}$ is the surface escape velocity of the object, $\tau \equiv 2 \Robj \sigma n_T$ is the optical depth,\footnote{Physically, this represents the average number of scatters a DM particle would undergo if it traversed the diameter of the object.} $p_N(\tau)$ is the probability a DM particle undergoes $N$ scatters as it traverses the object, and $g_N(u)$ is the probability a DM particle with initial speed $u$ becomes gravitationally bound having scattered $N$ times. One can show that for $p_N(\tau)$ we have~\cite{Ilie:2020Comment}: 
\begin{equation}
    p_N(\tau) = \frac{2}{\tau^2}\left(N+1 - \frac{\Gamma(N+2, \tau)}{N!}\right).
\end{equation}
Further progress can be made on this expression if we assume a particular form of $f(u)$. As is standard for thermal distributions, which is expected for DM particles in the halos hosting exoplanets, $f(u)$ is taken to be a Maxwell-Boltzmann distribution, and the partial capture rate is thus:
\begin{equation}
    C_{N}=\frac{1}{3}\pi \Robj^{2} p_{N}(\tau) \frac{\sqrt{6} n_{X}}{\sqrt{\pi} \bar{v}}\left(\left(2 \bar{v}^{2}+3 \vesc^{2}\right)-\left(2 \bar{v}^{2}+3 v_{N}^{2}\right) \exp \left(-\frac{3\left(v_{N}^{2}-\vesc^{2}\right)}{2 \bar{v}^{2}}\right)\right),
    \label{eq:CN_MBDistribution}
\end{equation}
where $\vbar$ is the DM velocity dispersion in the capturing region of the halo, and:
\begin{equation}
    v_N^2 \equiv \vesc^2\left(1 - \langle z\rangle\beta_+ \right)^{-N},
\end{equation}
where $\langle z\rangle \in [0,1]$ is the average of a kinematic parameter $z$, defined as follows: a DM particle incident on the object will have an initial kinetic energy $E_0
= \frac{1}{2} m_X w^2$. After elastically colliding with a nucleus, the particle will lose an amount of energy
$\Delta E = z \beta_+ E_0$, where $z$ is related the scattering angle in the center of mass frame by $z=\sin^2(\theta_{\text{CM}}/2)$ \cite{Dasgupta:2019juq}, and $\beta_+ \equiv \frac{4 m m_X}{(m + m_X)^2}$.

In~\cite{Ilie:2020Comment,Ilie:2020PopIII}, analytic expressions for the total capture rate were found and verified for the limiting regime of $\vesc \gg \vbar$ and $\langle z \rangle \sim \frac{1}{2}$. The result, nicely summarized by Fig.~4 of \cite{Ilie:2020PopIII}, is a set of analytic expressions for different regions of $\sigma-m_X$ parameter space, divided according to the optical depth, $\tau$, and the dimensionless kinematic variable $k \equiv \frac{3 \vesc^2}{\vbar^2} \frac{\min(m_X, m)}{\max(m_X, m)}$. In this work, we apply a more general analysis and acquire analytic capture rates regardless of $\vbar-\vesc$ hierarchy, and for all possible values of $\langle z\rangle$. This was motivated by the fact that many of the exoplanet models considered stradle the boundary of $\vbar\sim \vesc$, so it was necessary to have a clear picture of the analytic capture rates in this more general scenario.

\begin{figure}[bht]
    \centering
    \includegraphics[width=1\textwidth]{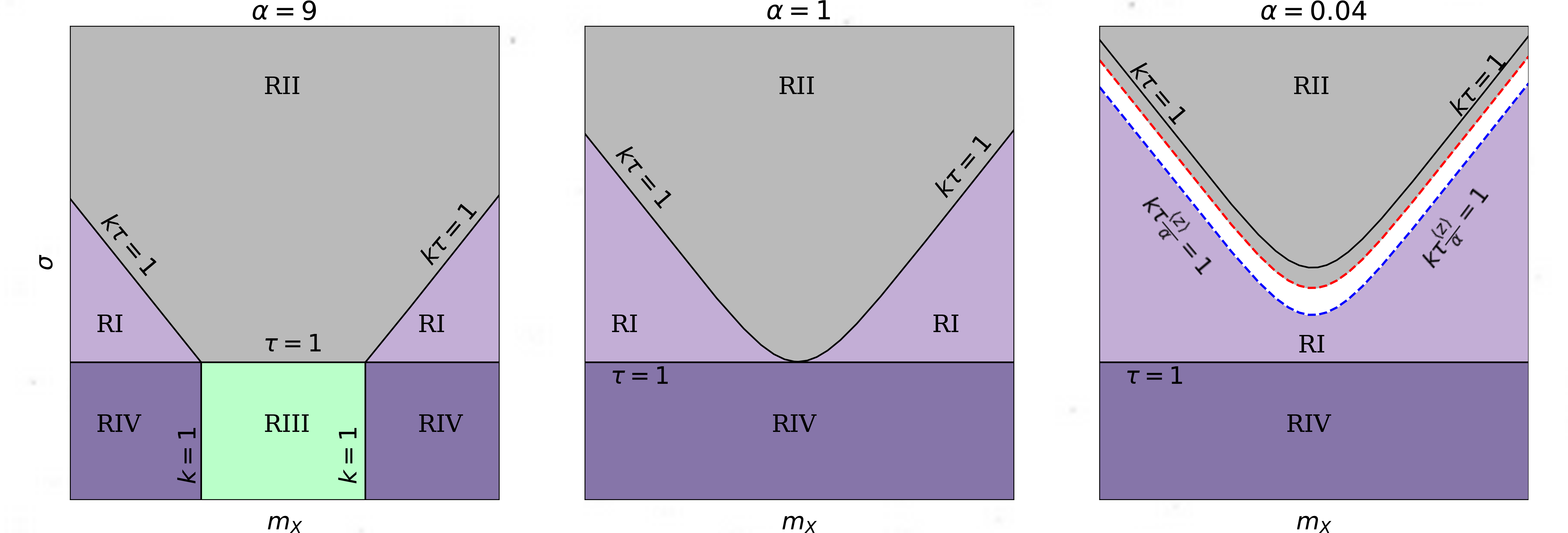}
    \caption{Schematic depiction of $\sigma-m_X$ parameter space indicating the regions associated with the analytic capture rates presented in Eqs.~(\ref{eq:CtotRI})-(\ref{eq:CtotRIV}). The first panel describes the results found in~\cite{Ilie:2020PopIII} where $\vesc > \vbar$, which corresponds to $\alpha >1$. When $\alpha \leq 1$, we see the erasure of Region III as seen in the central and right panels. In the right panel, when $\alpha < 1$, we see an interesting phenomenon occur: the boundary between Region I (blue dashed line) and Region II (red dashed line) become disassociated. The reason for this is that the approximation leading to the emergence of Region I breaks down before the capture rate becomes saturated in Region II. In the white region of the right panel, between the boundaries of Region I and Region II, the capture rate can be calculated numerically but we are unable to find a simple closed-form solution (although in~\cite{Ilie:2023Capture} we found an analytic--albeit very cumbersome--closed form for the entire $\sigma-m_X$ parameter space). Intriguingly, the boundary of Region II is also shifted below the $k\tau = 1$ line, though finding its exact location can only be done numerically at this point. However, we can say for certain that it occurs somewhere between the $k\tau = 1$ line and the Region I boundary, given now by the following condition: $k\tau\frac{\langle z\rangle}{\alpha}=1$. }
    \label{fig:AnalyticCapture_Schematic}
\end{figure}

\begin{figure}[bht]
    \centering
    \includegraphics[width=1\textwidth]{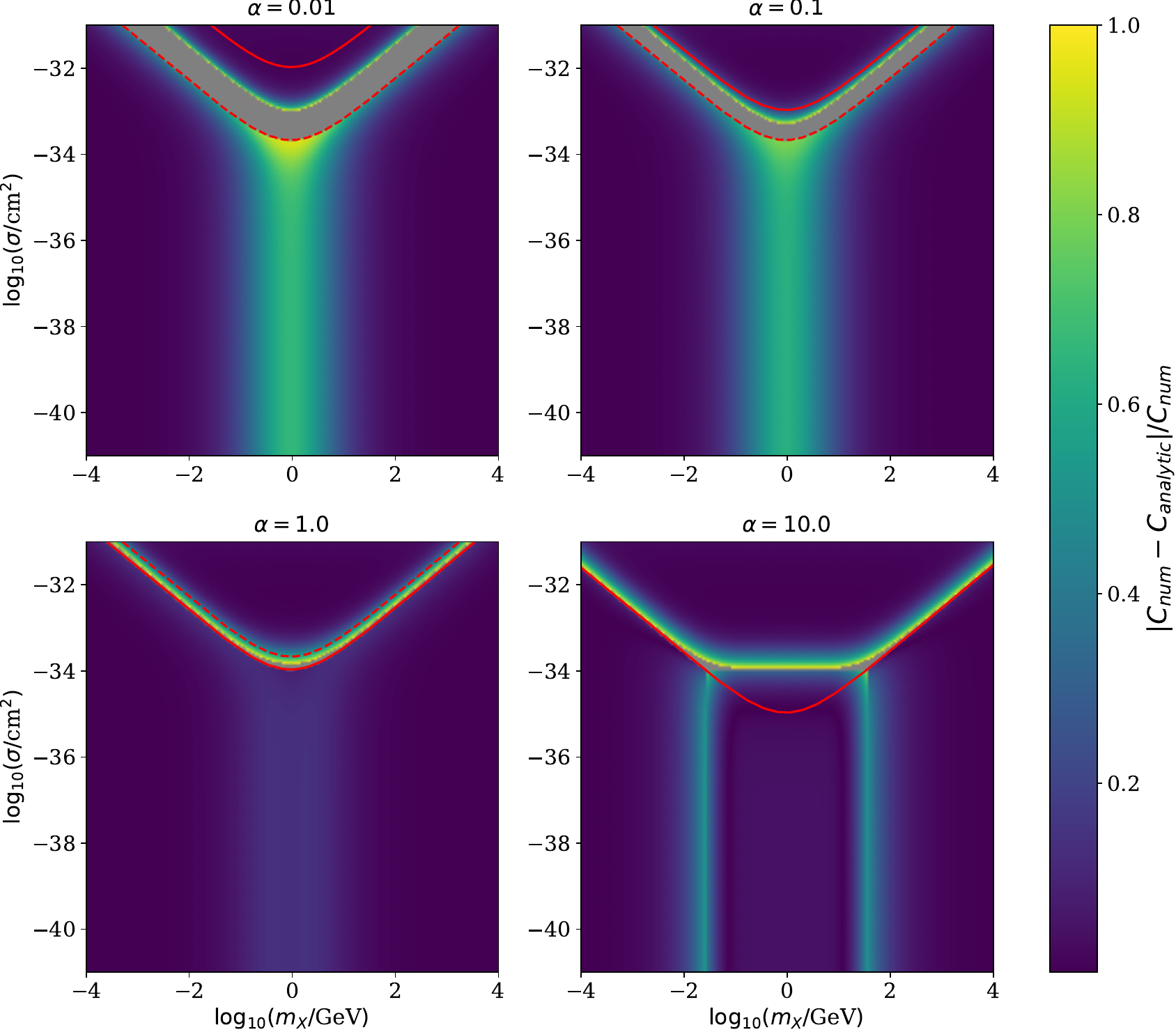}
    \caption{Relative error in the analytic approximations of the total capture rate given by Eqs.~(\ref{eq:CtotRI})-(\ref{eq:CtotRIV}) as compared to the numerical capture rate in Eq.~\ref{eq:CtotBasic} and Eq.~\ref{eq:CN_MBDistribution}. Parameters are chosen conveniently to represent different values of $\alpha$. In all panels, the solid red line represents $k\tau=1$, while the dashed red lines represent $k\tau \frac{\langle z\rangle}{\alpha}=1$. In the top panels, when $\alpha < 1$, the gray region represents the decoupling between regions I and II. We clearly see that the boundary of Region II, which is the upper boundary of the grey region, lies below the $k\tau=1$ line and above the Region I boundary, which is the lower boundary of the grey region. This contrasts the lower panels, when $\alpha \geq 1$, where the boundaries of Region I and II coincide at the solid red line, i.e. $k\tau=1$.}
    \label{fig:Capture_AnalyticvsNumerical}
\end{figure}

We summarize the results below, and derive them fully in Appendix~\ref{sec:AnalyticCapture}. First, it is convenient to define a couple of parameters:
\begin{equation}
    \alpha \equiv \frac{3 \vesc^2}{2\vbar^2} \langle z\rangle,
    \label{eq:alpha_def}
\end{equation}
and
\begin{equation}
    k \equiv \alpha \beta_+.
    \label{eq:kDef}
\end{equation}
Then, two distinct pictures emerge when $\alpha \geq 1$ and $\alpha < 1$. The case of $\alpha \gg 1$ was addressed in~\cite{Ilie:2020PopIII} where we found that the parameter space naturally divided into 4 regions according to the values of $\tau$, $k$, and $k\times\tau$, when compared to unity. These expressions are valid not only in the $\alpha \geq 1$ regime, as a very similar, albeit slightly different, picture emerges in the $\alpha < 1$ regime. The four regions found in~\cite{Ilie:2020PopIII} thus completely describe the analytic capture rates in most of the $\sigma-m_X$ parameter space, though their distributions throughout the space depend on $\alpha$. The analytic capture rates in the four regions identified by us (see Fig.~\ref{fig:AnalyticCapture_Schematic}) are:
\begin{equation}
    C = 2 A k\tau \vesc^2, \quad\quad \text{(Region I)},
    \label{eq:CtotRI}
\end{equation}
\begin{equation}
    C = A \left(2 \vbar^2 + 3\vesc^2\right), \quad \text{(Region II)},
    \label{eq:CtotRII}
\end{equation}
\begin{equation}
    C = 2 A \tau \vesc^2, \quad\quad \text{(Region III)},
    \label{eq:CtotRIII}
\end{equation}
\begin{equation}
    C = 2 A k\tau \vesc^2, \quad\quad \text{(Region IV)},
    \label{eq:CtotRIV}
\end{equation}
where $A\equiv \frac{1}{3}\pi \Robj^2 \sqrt{\frac{6}{\pi}} \frac{n_X}{\vbar}$. The simple capture rates in these regions stem from various approximations, outlined in Appendix~\ref{sec:AnalyticCapture}. The key differences in the $\alpha < 1$ regime, as one can see from the schematic depiction in Fig.~\ref{fig:AnalyticCapture_Schematic}, is the erasure of Region III, the decoupling of the Region I and Region II boundary, and the emergence of a region between Region I and Region II where either approximation breaks down. When $\alpha \geq 1$, the boundary of Region II, which coincides with the boundary of Region I, is well-defined by the $k\tau=1$ line, a fact which no longer holds in the $\alpha < 1$ case. The new boundary for Region I is given by the line $\frac{\langle z\rangle k\tau}{\alpha} = 1$. However, as of now we could not find an analytic form for the boundary of Region II in this scenario, though we confirmed numerically that it lies between $k\tau=1$ and $\frac{\langle z\rangle k\tau}{\alpha} = 1$. Another very intriguing feature of these equations, which is evidenced in Fig.~\ref{fig:AnalyticCapture_Schematic} and Fig.~\ref{fig:Capture_AnalyticvsNumerical} by the smooth transition between Region I and IV, is the degeneracy of the capture rate in Regions I and IV. This is a rather counter-intuitive result since, in deriving each result, we assume opposite limits for the optical depth, $\tau$, i.e. $\tau\gg1$ for Region I , and $\tau\ll1$ for Region IV.

In Fig.~\ref{fig:AnalyticCapture_Schematic}, we plot the $\sigma-m_X$ parameter space schematically, indicating the regions identified in Eqs.~(\ref{eq:CtotRI})-(\ref{eq:CtotRIV}). In this diagram, we see the role that the $\vbar-\vesc$ hierarchy plays in determining the location of the regions, and therefore where the different analytic capture rates are valid. When $\alpha > 1$, we see the already established picture emerge as shown in~\cite{Ilie:2020PopIII}: four regions of parameter space, divided according to the boundaries of $\tau = 1$, $k = 1$, and $k\tau = 1$. When $\alpha \leq 1$, we see the disappearance of Region III since $k \leq 1$ always in this limit. We also begin to see the splitting of boundaries between Region I and Region II, as the Region I approximations break down before the capture rate is saturated. The new boundary for RI is determined by the line $\langle z\rangle k \tau/\alpha = 1$ as is shown in Appendix~\ref{sec:AnalyticCapture}. In between these boundaries we have yet to produce a simplified analytic useful closed form for the capture rate, though one can still compute it numerically. However, we point out that in~\cite{Ilie:2023Capture} we have obtained a full closed form (albeit very cumbersome) of the total capture rates covering the entirety of the $\sigma-m_X$ parameter space, for any $\vesc$ and $\vbar$. Lastly, we point out that the boundary of Region II also appears lower than the $k\tau = 1$ line, another difference when in the $\vbar > \vesc$ regime.

In Fig.~\ref{fig:Capture_AnalyticvsNumerical}, we validate our analytic approximations for the DM capture rates in the four distinct regions (RI-RIV) identified. Specifically, we plot the relative error in the analytic capture rates of Eqs.~(\ref{eq:CtotRI})-(\ref{eq:CtotRIV}) as compared to the numerical capture rate (Eq.~\ref{eq:CtotBasic} combined with Eq.~\ref{eq:CN_MBDistribution}). Each panel corresponds to a different value of $\alpha$, as labeled in the plot. In this figure, we see the incredible accuracy in the approximations of the capture rate, as well as the emergence of the regions depicted schematically in Fig.~\ref{fig:AnalyticCapture_Schematic}. Also, as mentioned above, we clearly see the degeneracy of our analytic estimates in comparing Regions I and IV which, remarkably, produce the exact same form for the analytic capture rate  despite requiring different assumptions in their derivations. Some key features that emerge when $\alpha \leq 1$ are the erasure of a Region III (which is clearly present when $\alpha > 1$) and the failure to approximate the capture rate well between Region I and Region II. As discussed above in reference to Fig.~\ref{fig:AnalyticCapture_Schematic}, we find that the transition out of Region I and the transition into Region II no longer coincide when $\alpha < 1$. More details on this are found in Appendix~\ref{sec:AnalyticCapture}.

\subsection{Evaporation}
Evaporation is the reverse process of capture. DM particles that are gravitationally bound will interact with the object's constituents and, through these interactions, it is possible for a particle to gain enough energy to escape the object. Like capture, evaporation was first studied in the context of the Sun, and has since been applied to a variety of astrophysical objects. The rate of evaporation per particle in an object is given by the following prescription~\cite{Garani:2022}:
\begin{equation}
    E = \frac{1}{N_X}\int_0^{\Robj} s_\text{evap}(r) n_X(r)~4\pi r^2~dr \int_0^{\vesc(r)} f_X(\boldsymbol{w}, r)~4\pi w^2~dw \int_{\vesc(r)}^\infty \mathcal{R}^+ (w\rightarrow v)~dv,
    \label{eq:Evaporation_IntegralForm}
\end{equation}
where $N_X$ is the total number of captured DM particles, $v$ represents the up-scattered speeds sufficient for escape, $\vesc(r)$ is the escape velocity at radius $r \leq \Robj$, $\boldsymbol{w}$ represents the velocity of DM particles with local distribution $f_X(\boldsymbol{w}, r)$ (taken to be a Maxwell-Boltzmann distribution in this paper), $\mathcal{R}^+(w\rightarrow v)$ is the rate of up-scattering from speed $w$ to $v$, $n_X(r)$ is the DM number density distribution within the object, and $s_\text{evap}(r)$ is a suppression factor which accounts for two distinct possibilities: 1.~DM leaving in non-radial trajectories, and  2.~DM down-scattering while on the way out. The suppression factor is given by~\cite{Garani:2022}:
\begin{equation}
    s(r) = e^{-\tau(r)} \underbrace{\frac{7}{10} \frac{1 - e^{-\frac{10}{7}\tau(r) }}{\tau(r)}}_{\text{Angular Factor}} \underbrace{{}_0 F_1\left(; 1 + \frac{2}{3} \hat{\phi}(r); \tau(r)\right)}_{\text{Down-scatter Factor}},
    \label{eq:SuppressionFactor}
\end{equation}
where ${}_0 F_1\left(;b;z\right)$ is the confluent hypergeometric limit function and:
$$\tau(r) \equiv \int_r^{\Robj} \sigma n(r') dr',~ \hat{\phi}(r) \equiv \frac{\frac{1}{2} m_X \vesc^2(r)}{T(r)}$$
Assuming a velocity-independent scattering cross section, the rate of upscattering can be shown to carry the following functional dependence:
\begin{equation}
    \mathcal{R}^+ (w\rightarrow v) = \mathcal{R}^+ (w,~v,~m,~m_X,~T(r),~n(r),~\sigma),
\end{equation}
where $n(r)$ is the hydrogen number density distribution and $T(r)$ is its temperature distribution. More details can be found in Appendix~A of~\cite{Garani:2017} for the full form as well as details on how to arrive there. 

To compute the evaporation rate, then, a number of features of the object and how DM distributes itself in the object are required. In the case where DM interacts very little with the object's constituents, the isothermal limit, its distribution is given by a rather straightforward isothermal distribution. However, if there are a large number of interactions between DM and the object's constituents, the local thermal equilibrium (LTE) limit, understanding the object's distribution properties is essential to understanding the DM distribution. In Section~\ref{sec:DMDistributions}, we discuss how to approach this problem for Jupiters and Brown Dwarfs, outlining how polytropes with index $n=1-1.5$ can be used to model these objects and how that information is further employed to determine the DM distribution in the high-interaction limit. Typically, one finds themselves in a regime of either very high or very low interaction and in these cases, computing the evaporation rate is relatively simple, since finding the profile is straightforward. However, when straddling the border between isothermal and LTE, some care must be taken. To tackle this problem, we follow~\cite{Garani:2017} and interpolate between the two regimes via a parameter called the Knudsen Number. It is defined as:
\begin{equation}
    \text{Kn} = \frac{l_{\text{mfp}}}{L},
    \label{eq:KnudsenNumber}
\end{equation}
where $l_{\text{mfp}}\equiv (\sigma n_T)^{-1}$ is the mean free path, $n_T$ is the average number density of hydrogen, and $L$ is the length scale of the system. Thus, when $\text{Kn} \gg 1$, the so-called ``Knudsen limit,'' DM interacts very little with the surrounding medium since $l_{\text{mfp}} \gg L$. Conversely, when $\text{Kn} \ll 1$, there is a high level of interaction. Determining the length scale of the system can be done by finding the radius $r_X$ at which the DM particles are in thermodynamic equilibrium with the core~\cite{Spergel:1985}. In other words, solving the following:
\begin{equation}
    \frac{3}{2} T_c = m_X \Phi(r_X),
\end{equation}
where $\Phi(r)$ is the gravitational potential of the object. The solution is given by~\cite{Spergel:1985}:
\begin{equation}
    L = r_X = \left(\frac{9}{4\pi}\frac{T_c}{G \rho_c m}\right)^{1/2} \sqrt{\frac{m}{m_X}}.
\end{equation}
Following ~\cite{Garani:2017}, we define the Knudsen transition function by:
\begin{equation}
    f(\Kn) = \frac{1}{1 + \left(\Kn/0.4\right)^2},
    \label{eq:KnudsenTransitionFunction}
\end{equation}
which is the interpolation function between the isothermal and the LTE limits. Note the following, which is consistent with this interpretation of $f(K)$ as an interpolation:
$$\lim_{\text{Kn}\to \infty} f(K) = 0,$$
$$\lim_{\text{Kn} \to 0} f(K) = 1.$$
With this,~\cite{Garani:2017} defines the product $n_X(r) f_X(\mathbf{w}, r)$ in the intermediary regime ($\text{Kn}\sim 1$) to be:\footnote{If one is interested in a more accurate approach to the Knudsen transition regime, see~\cite{Banks_2022}, where Monte Carlo simulations were carried out to model energy transport of DM in astrophysical bodies.}
\begin{equation}
    n_X(r) f_X(\mathbf{w}, r) = f(\text{Kn}) n_{X}^{\text{LTE}} f_X(\mathbf{w}, r)^\text{LTE} + \left[1-f(\text{Kn})\right] n_{X}^{\text{ISO}} f_X(\mathbf{w}, r)^\text{ISO}.
    \label{eq:nX_fX_interpolation}
\end{equation}
By integrating over the DM velocity space ($\mathbf{w}$), one gets:
\begin{equation}
    n_X(r) = f(\text{Kn}) n_{X}^{\text{LTE}} + \left[1-f(\text{Kn})\right]n_{X}^{\text{ISO}}.
    \label{eq:nx_r_Transition}
\end{equation}
Combining Eqs.~\ref{eq:Evaporation_IntegralForm} and \ref{eq:nX_fX_interpolation} and noting that $f(K)$ has no dependence on the variables being integrated, we see that we can generally write the evaporation rate as:
\begin{equation}
    E = f(\text{Kn}) E_{\text{LTE}} + \left[1-f(\text{Kn})\right] E_{\text{ISO}},
    \label{eq:Evaporation_interpolated}
\end{equation}
where $E_{\text{LTE}}$ and $E_{\text{ISO}}$ are given by Eq.~\ref{eq:Evaporation_IntegralForm} where the distributions $n_X$ and $f_X$ are taken in the LTE/isothermal limits. Note here that, in both the LTE and isothermal case, we do not need to actually know the total number of DM particles in the object, $N_X$, despite its appearance in Eq.~\ref{eq:Evaporation_IntegralForm}. This is because we have:
$$E \sim \frac{\int  n_X(r) g(r) dV}{N_X} \sim \frac{n_X(r=0) \int \frac{n_X(r)}{n_X(r=0)} g(r) dV}{ n_X(r=0) \int \frac{n_X}{n_X(r=0)} dV} \sim \frac{\int \frac{n_X}{n_X(r=0)} g(r)  dV}{ \int \frac{n_X}{n_X(r=0)} dV},$$
where $g(r)$ represents the rest of whats being integrated in Eq.~\ref{eq:Evaporation_IntegralForm}, and $n_X/n_X(r=0)$ is the dimensionless DM distribution, whose form we know exactly in the isothermal and LTE regimes, as discussed in Section~\ref{sec:DMDistributions}. One must take care when applying the interpolating prescription in Eq.~\ref{eq:Evaporation_interpolated}, however, since you may enter certain regimes where it is no longer valid. This is as a result of the exponential cutoff of the evaporation rate above the so-called ``evaporation mass.'' For example, if $\text{Kn} \ll 1$ so that we should have $E = E_{\text{LTE}}$ but whenever, for a given set of parameters,  $E_{\text{LTE}}$ is exponentially suppressed more than $E_{\text{ISO}}$, one will find $E\approx \left[1 - f(\text{Kn})\right] E_{\text{ISO}}$. This is because the first term would be more exponentially suppressed than the second term and the $\left[1-f(\text{Kn})\right]$ factor does not send $E_{\text{ISO}}$ to $0$ quickly enough. To avoid this issue, we stipulate a condition when computing the evaporation rate that when one is well into the $\text{Kn}\ll 1$ or $\text{Kn}\gg 1$ regime, we switch to $E = E_\text{LTE}$ or $E_\text{ISO}$ respectively. 

\subsection{Annihilation}

Once DM is captured, if it is self-annihilating it could contribute a heat source to the object through model-dependent annihilation channels. As mentioned in the beginning of the section, we consider the CoSIMP DM model for the purpose of obtaining sensitivity limits in the sub-GeV mass region, where, as suggested by~\cite{Leane:2020wob}, exoplanets should excel as DM detectors. For future reference we consider here other $2\to 2$ Dark Matter models, including WIMPs and Superheavy DM models, in approximating annihilation coefficients for heavy ($m_X \gg 1\unit{GeV}$) DM. The CoSIMP model assumes a strong coupling between DM and SM particles, and the thermally averaged annihilation cross sections, denoted by $\langle\sigma v^2\rangle$, are constrained to produce the correct relic abundance of DM in the universe~\cite{Smirnov:2020}. We focus on CoSIMP annihilation over SIMP annihilation since, while these number changing ($3\to2$) process can happen simultaneously~\cite{Hochberg:2014}, in exoplanets the CoSIMP model dominates the annihilation process due to the high density of SM particles in exoplanets that aid the efficiency of annihilation~\cite{Leane:2020wob}. In natural units, the annihilation interaction cross-section that gives the correct relic abundance is~\citep{Leane:2020wob}:
\begin{equation}\label{eq:sv2}
    \langle\sigma v^2\rangle = 10^3 \left(\frac{1\unit{GeV}}{m_X}\right)^3 \unit{GeV}^{-5}.
\end{equation}
For $2\to 2$ models outside of the WIMP window, one could consider thermally averaged cross sections $\langle\sigma v\rangle$ bounded by the Griest-Kamionkowski bound~\cite{Griest:1990}:
\begin{equation}
    \langle\sigma v\rangle = \frac{4\pi}{m_X^2 v_X},
\end{equation}
or, in the WIMP regime, the ``WIMP miracle'' thermally averaged cross section:
\begin{equation}
    \langle \sigma v\rangle_\text{WIMP} \simeq 10^{-23}\unit{cm^3 s^{-1}}. 
\end{equation}
With this in mind, the $N_X$-independent annihilation coefficient for the CoSIMP model is~\cite{Leane:2020wob}:
\begin{equation}
    C_A = \frac{\int dV~n_X^2 n \langle\sigma v^2\rangle}{\left(\int dV n_X\right)^2},
    \label{eq:Ca_CoSIMP}
\end{equation}
while for the $2\to 2$ models it is:
\begin{equation}
    C_A = \frac{\int dV~n_X^2 \langle\sigma v\rangle}{\left(\int dV n_X\right)^2}.
    \label{eq:Ca_2to2}
\end{equation}
If we take Eq.~\ref{eq:Ca_CoSIMP}, for example, and rewrite it in the following way:
\begin{equation}
    \Gamma = C_A N_X^2 = \int dV~n_X^2 n \langle\sigma v^2\rangle,
\end{equation}
we see that it is just an explicit way to calculate the total annihilation rate by summing the local annihilation rates $d\Gamma = \langle\sigma v^2\rangle n_X^2 n~dV $ over the entire object. Note that, given a DM distribution of the form $n_X(r) = n_X(0) \eta(r)$, for some $\eta(r)$ such that $\eta(0)=1$, that the dimensional component of the distribution cancels between numerator and denominator of Eq.~\ref{eq:Ca_CoSIMP}, so only functional form $\eta(r)$ of the distribution is necessary. This is the benefit of considering an annihilation coefficient $C_A$: it gives an insight into the annihilation efficiency of DM in the object without needing to know the number of DM particles present. Furthermore, it is the annihilation coefficient, along with the capture rate and the per-particle evaporation rate, that determine the time-scale at which equilibrium between these processes occur, as per Eq.~\ref{eq:dNx_DiffEq}. In section~\ref{sec:DMDistributions} we find the functional form of $\eta(r)$ and in Appendix~\ref{sec:analyticApproximations} we use this result to analytically approximate $C_A$ in the high DM mass limit for $2\to 2$ DM models. 

\section{Constraining Dark Matter with exoplanets}

\subsection{Background on exoplanets and exoplanet searches}

When considering the promise of astrophysical objects as probes of DM through the processes of capture and subsequent annihilation, there are a number of desirable features that will tend to make a DM signal stronger. Ideally, one would like an object with a high baryonic density, large surface area, small internal heat sources, and in a dense DM environment. However, many times only one or a few of the conditions listed above are met. For instance, as the universe expands, the density of DM in the universe dilutes, so objects in early DM halos, such as Population III stars, have the advantage of being in very dense DM environments. From an observational point of view, however, such objects pose a challenge due to their large distance. On the other hand, exoplanets  in the Milky Way can provide nearby  astrophysical objects with little to no internal heat sources, and as such they can be used as probes of DM, as shown initially by~\cite{Leane:2020wob}.

Exoplanets are planetary bodies lying outside of our solar system. These bodies can take a wide variety of forms, and may be broadly categorized into ``Earth-like'' planets, Jupiters, and Brown Dwarfs based on their size and make-up. Earth-like planets are characterized by their relatively small size and rocky interiors. Jupiters are composed of gas, predominantly hydrogen, and have radii of approximately Jupiter's radius and range in mass from the mass of Jupiter to about 14 times the mass of Jupiter. Brown Dwarfs are the largest of the three ranging in mass from about $14 M_J$ to $75 M_J$ and straddle the planet-star boundary. Remarkably, Brown Dwarfs have approximately the same radii as Jupiters, making them much denser objects and ideal as DM probes. However, with this increased density comes additional internal heating and longer cooling timescales that may pose a challenge in identifying a DM signal. Because of their larger size and density, we focus on Jupiters and Brown Dwarfs as DM probes in this work. To give a sense of exoplanet abundance in our own Milky Way, note that microlensing observations have been used to estimate that every star in our galaxy has, on average, at least one bound exoplanet~\cite{Cassan_2012}, and the number of stars in the Milky Way is of order $\sim 100$ billion. Furthermore, ~\cite{Cassan_2012} find that about $20\%$ of stars in the Milky Way have a Jupiter to Brown Dwarf sized companion planet. Exoplanets are ideal probes of DM primarily due to their proximity. Searches targeting exoplanets within our own galaxy of course have the advantage of location. To date, there have been a number of exoplanet detections within our own galaxy~\cite{ExoplanetList}, with a few recent hot gas giants being detected in one single month alone~\cite{Currie:2023, gupta2023higheccentricity}, making these promising candidates as probes of DM. However, the candidate HIP 99770 b is at most 414Myr in age~\citep{Currie:2023}, which accounts for its relatively high temperature. Although the candidate TOI 4727 b in \cite{gupta2023higheccentricity} is more than 1Gyr in age, its temperature is typical for exoplanets at its orbital distance assuming radiative properties similar to Jupiter. While those two examples (TOI 4727 b HIP 99770 b) of hot gas giants did not prove to be particularly useful as DM probes, the promise of future detections of rogue overheated exoplanets with dedicated exoplanet surveys is high. Moreover, with its excellent field of view,  the upcoming Roman Space Telescope~\cite{Johnson_2020} adds to the prospects of discovering numerous objects that can serve as Dark Matter probes, such as those considered in this paper: rogue overheated Jupiters and Brown Dwarfs. 

\subsection{Particle distributions in exoplanets}\label{sec:DMDistributions}
In this section, we outline how we go about approximating the hydrogen and DM distribution in Jupiters and Brown Dwarfs. We start by describing distributions of hydrogen before discussing how we find the DM distributions in these objects. In Appendix~\ref{sec:analyticApproximations}, we use the results of this section to approximate the annihilation coefficient for highly-cored profiles. It is worth emphasizing here that, as done in~\cite{Leane:2020wob} and the majority of works studying the astrophysical observables of annihilating DM, we are making the simplifying assumption that the effects of DM heating on the object does not change its structure. Going beyond this commonly used approximation  would require incorporating the effects of DM on the heating and energy transport in stellar evolution codes, such as \texttt{MESA}~\cite{Paxton2011}.

Jupiters and Brown Dwarfs are made predominantly of hydrogen and, to a good approximation, can be modelled by polytropes of with an index ranging from $n=1$ to $n=\frac{3}{2}$. We follow~\cite{Garani:2022} in our choices of polytrope index, who chose $n=1$ for Jupiters by comparing to the expected models of gas giants in our solar system~\cite{Helled_2010, MARLEY2014743, Stevenson} and $n=3/2$ for Brown Dwarfs, as they are dominated by convective energy transport~\citep{Chabrier_2000}. The $n=1$ polytrope has a closed-form solution while the $n=1.5$ must be solved numerically. Polytropic distributions of index $n$ are of the form $\rho(\xi) = \rho_c \theta^n(\xi)$, where $\xi$ is a dimensionless radial variable and $\theta(\xi)$ is a solution to the Lane-Emden equation:
\begin{equation}
    \frac{1}{\xi^2}\frac{d}{d\xi}\left(\xi^2 \frac{d\theta}{d\xi}\right) = -\theta^n.
    \label{eq:LaneEmden}
\end{equation}
The above equation is a restatement of the condition of hydrostatic equilibrium of a self-gravitating, spherically symmetric fluid. The boundary conditions are given by: $\theta(0) = 1$ and $\theta'(0)=0$. The dimensionless radial parameter is defined by $\xi = r \frac{\xi_1}{\Robj}$, where $\xi_1$ is the first positive solution of $\theta(\xi) = 0$. Thus, $\xi = \xi_1 \iff r=\Robj$. If we assume the hydrogen behaves like an ideal gas, as is the case for Jupiters, the temperature profile of the object is given by: $T(\xi) = T_c \theta(\xi)$. Remarkably,  the same relationship is also valid the interior of Brown Dwarfs~\cite{Auddy:2016}.

Furthermore, one can show that for a general polytrope, the mass enclosed by a radius $\xi$ is: 
\begin{equation}
M(\xi)=-4\pi\rho_c\left(\frac{\Robj}{\xi_1}\right)^3\xi^2\frac{d\theta}{d\xi}.
\end{equation}
If the minus sign is concerning, note that $\theta'(\xi) \leq 0$ for all $\xi\in \left[0, \xi_1\right]$. The gravitational potential is defined in the usual way:

\be
\Phi(r)\equiv \int_0^r dr'\frac{G M(r')}{r'^2}.
\ee
Using the dimensionless variable $\xi$, as appropriate for polytropes, this becomes: 

\begin{equation}
    \Phi(\xi) \equiv \frac{\xi_1}{R}\int_0^\xi d\xi' \frac{G M(\xi')}{\xi'^2}.
\end{equation}
Then, we can compute this in terms of the solution to the Lane-Emden equation as:
\begin{equation}
    \Phi(\xi) = \frac{1}{2}\frac{\xi_1}{\alpha_n}v_{esc}^2[1-\theta(\xi)],
\label{eq:PhiXi_Polytrope}
\end{equation}
where $\alpha_n \equiv \xi_1^2 \left( \frac{d\theta}{d\xi}\Bigr|_{\xi_1}\right)^2$. We note here that for the case of n=1, $\xi_1 = \alpha_n = \pi$. As we shall see later on, having an expression for the gravitational potential in terms of the Lane-Emden function will help us tremendously in computing $n_X(r)$ in the limit of high interactions.

For index $n=1$, Eq.~\ref{eq:LaneEmden}, along with the boundary conditions, has the following closed form solution:
\begin{equation}
    \theta(\xi) = \frac{\sin(\xi)}{\xi}
\end{equation}
Clearly, then, $\xi_1 = \pi$ for $n=1$. One can approximate $\theta(\xi)$ quite straightforwardly using Taylor expansion around $\xi = 0$:
\begin{equation}
    \theta(\xi) \simeq 1 - \frac{1}{6} \xi^2 + \frac{1}{120} \xi^4 - \frac{1}{5040} \xi^6+\mathcal{O}(\xi^8).
    \label{eq:thetaXi_approx}
\end{equation}
Note, however, that in this approximation, the first zero of $\theta(\xi)$ is at $\xi_1 \approx 3.08$, which one must keep in mind if integrating radially. In Appendix~\ref{sec:analyticApproximations}, which focuses on approximating integrals of the form:
$$\int dV n_X(\xi),$$
for very heavy DM, we will utilize this expansion but keep only the first two terms, which is actually a valid expansion for all index $n$ polytropes~\citep{Chandrasekhar1939}. In addition to being general (i.e. $n$ independent), the reason for truncating at the second order term is that the DM distribution profile $n_X$ becomes highly cored, whenever $m_X\gg 1$~GeV. Later in this section, we show how the Lane-Emden function appears in integrals over the DM distribution (See Eqs.~\ref{eq:nx_ISO_scaling}, \ref{eq:nX-xi_LTE}). Equation~\ref{eq:thetaXi_approx} is a sufficient, in most cases. In fact, without this approximation, computing such integrals for high DM mass numerically is challenging because of how cored the profiles are. For sub-GeV DM, however, we do not make any approximations on the Lane-Emden function since these integrals are straightforward to compute numerically.

\begin{figure}[bht]
    \centering
    \includegraphics[width=1\textwidth]{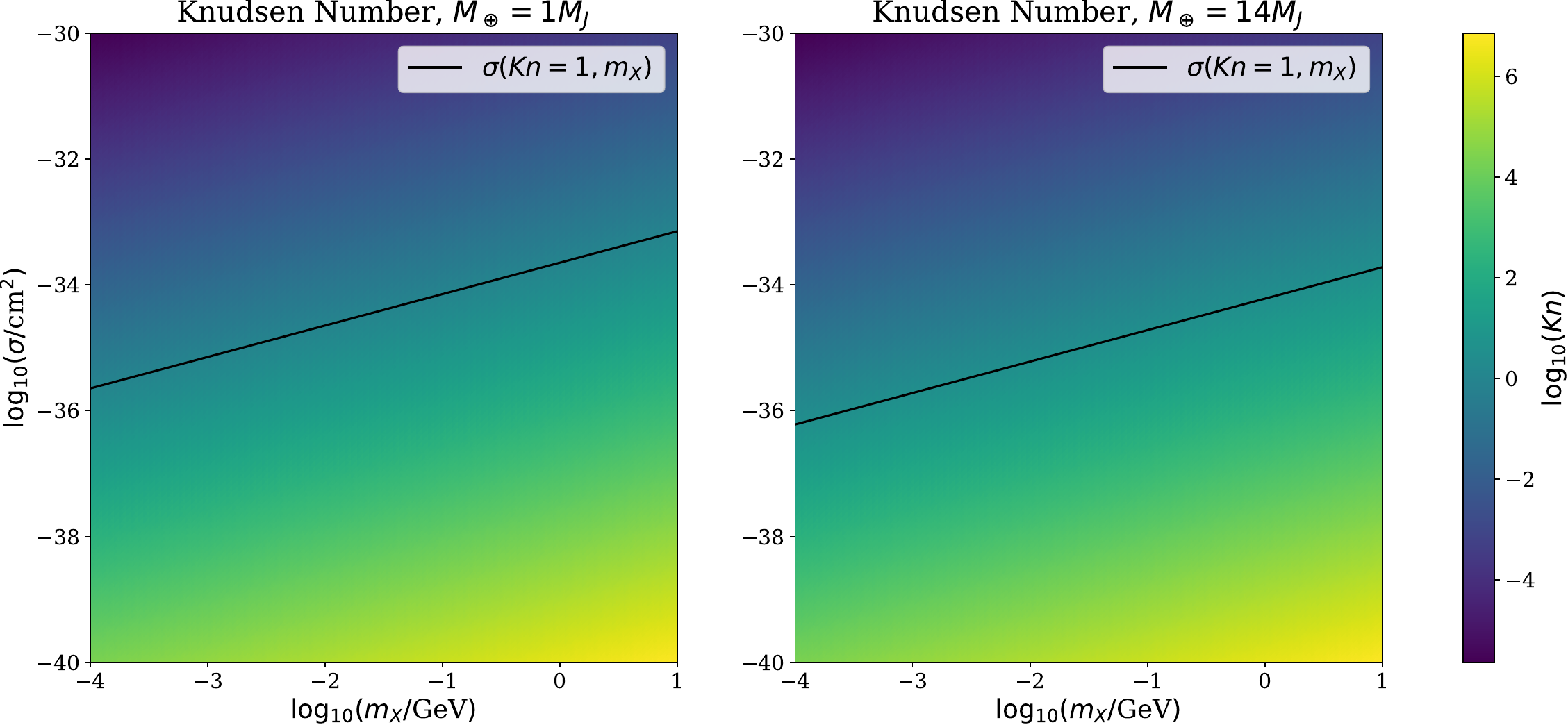}
    \caption{Knudsen number for a $\Mobj = 1 M_J$ (left panel) and $\Mobj = 14 M_J$ (right panel) jupiter-like exoplanet as a function of the DM-hydrogen scattering cross section ($\sigma$) and DM mass ($m_X$). The solid black line represents the $\text{Kn} = 1$ contour, and so we see that the region of $\sigma-m_X$ parameter space in which we are interested straddles the $\text{Kn}\sim 1$ boundary.}
    \label{fig:KnudsenNumber_sigma-mx}
\end{figure}

With the hydrogen distribution in hand, we turn to approximating the distribution of captured DM in these objects. We must first determine how frequently the DM particles interact with the hydrogen distribution. To do so, we use the Knudsen number, as defined in Eq.~\ref{eq:KnudsenNumber}. In Fig.~\ref{fig:KnudsenNumber_sigma-mx}, we scan over a part of the $\sigma-m_X$ parameter space for $\Mobj = 1 M_J$ and $\Mobj = 14 M_J$ Jupiters and compute the Knudsen number. In this figure, it is clear that, depending on the object, there is a region where we are not entirely in the $\Kn \gg 1$ or $\Kn \ll 1$ limit and so we must interpolate between the two regimes as we traverse the $\Kn\sim 1$ boundary using the interpolation function defined in Eq.~\ref{eq:KnudsenTransitionFunction}. Moreover, we note that this transition happens within the region of the $\sigma$ vs. $m_X$ parameter space probed by Jupiters (see Fig.~3 of~\cite{Leane:2020wob} and Fig.~\ref{fig:SensitivityLimits} in this paper).

Let us first suppose that we are in the Knudsen limit ($\text{Kn} \gg 1$). Then, as shown in \cite{Spergel:1985}, the DM distributes itself isothermally at a temperature $T_X$. The resulting profile is given by:
\begin{equation}
    n_X(\xi) = n_X(0) e^{-\frac{m_X \Phi(\xi)}{T_X}}.
    \label{eq:nX-xi_ISO}
\end{equation}
The main question to address in an isothermal profile is the temperature of the captured DM distribution. Following~\cite{Spergel:1985}, we find the distribution's temperature by solving the first energy moment of the collisional Boltzmann equation. In doing so, we are designating by $T_X$ the average DM temperature over a single orbit since, in reality, as DM particles orbit within the object, their temperature will undergo slight changes as they interact with the local medium. For time-independent distributions, such as the one we are considering, requiring that the first moment is satisfied is equivalent to there being no net flow of energy into the dark matter distribution from the surrounding hydrogen in the planet. The equation necessary to solve the DM temperature is~\cite{Spergel:1985}:
\begin{equation}
\int_{0}^{\xi_{1}} \theta(\xi) \exp \left(-\frac{\mu}{\Theta}\Phi(\xi)\right)\left(\frac{\Theta+\mu \theta(\xi)}{\mu}\right)^{1 / 2}[\Theta-\theta(\xi)] \xi^{2} d \xi=0,
\label{eq:BoltzmannFirstMoment}
\end{equation}
where we define the following dimensionless variables for computational convenience:
\begin{equation}
   \mu \equiv \frac{m_X}{m} \text{ , } \tilde{\Phi}(r) \equiv \frac{m \Phi(r)}{T_c} \text{ and } \Theta \equiv \frac{T_X}{T_c}.
\end{equation}
Combining Eq.~\ref{eq:nX-xi_ISO} and Eq.~\ref{eq:PhiXi_Polytrope}, we see that the dependence of $n_X(\xi)$ in the isothermal limit is given by: 
\begin{equation}
n_X \sim \exp\left\{-\gamma \left(1-\theta(\xi)\right)\right\},
\label{eq:nx_ISO_scaling}
\end{equation}
where $\gamma\sim m_X/T_X$. Additionally, one can show that if $m_X \gg m$, $T_X \approx T_c$ since in Eq.~\ref{eq:BoltzmannFirstMoment}, if $\mu\to\infty$, the exponential term cuts off the integral to the area around $\xi\to0$. Thus, the only way the LHS of Eq.~\ref{eq:BoltzmannFirstMoment} is zero is if $\Theta\to\theta(1)=1$, i.e. $T_X\to T_c$. In Appendix~\ref{sec:analyticApproximations}, we will explore the $m_X\to\infty$ limit to find approximations for various integrals involving $n_X$. This was done primarily by the recognition that numerical integration on exponential functions with large decay constants is challenging. In the opposite limit of $m_X\to 0$, one can show that $T_X \approx \text{const.}$, and so $\gamma\to 0$. In this case, the isothermal profile simply becomes constant, $n_X \simeq n_X(0)$. This is a fact that is unique to the isothermal distribution. As we will show below, the DM profile in the opposite ($\Kn \ll 1$) limit is not flat if $m_X\to 0$, but rather obtains a ``floating'' profile with higher densities toward the edge of the object.

\begin{figure}[bht]
    \centering
    \includegraphics[width=1\textwidth]{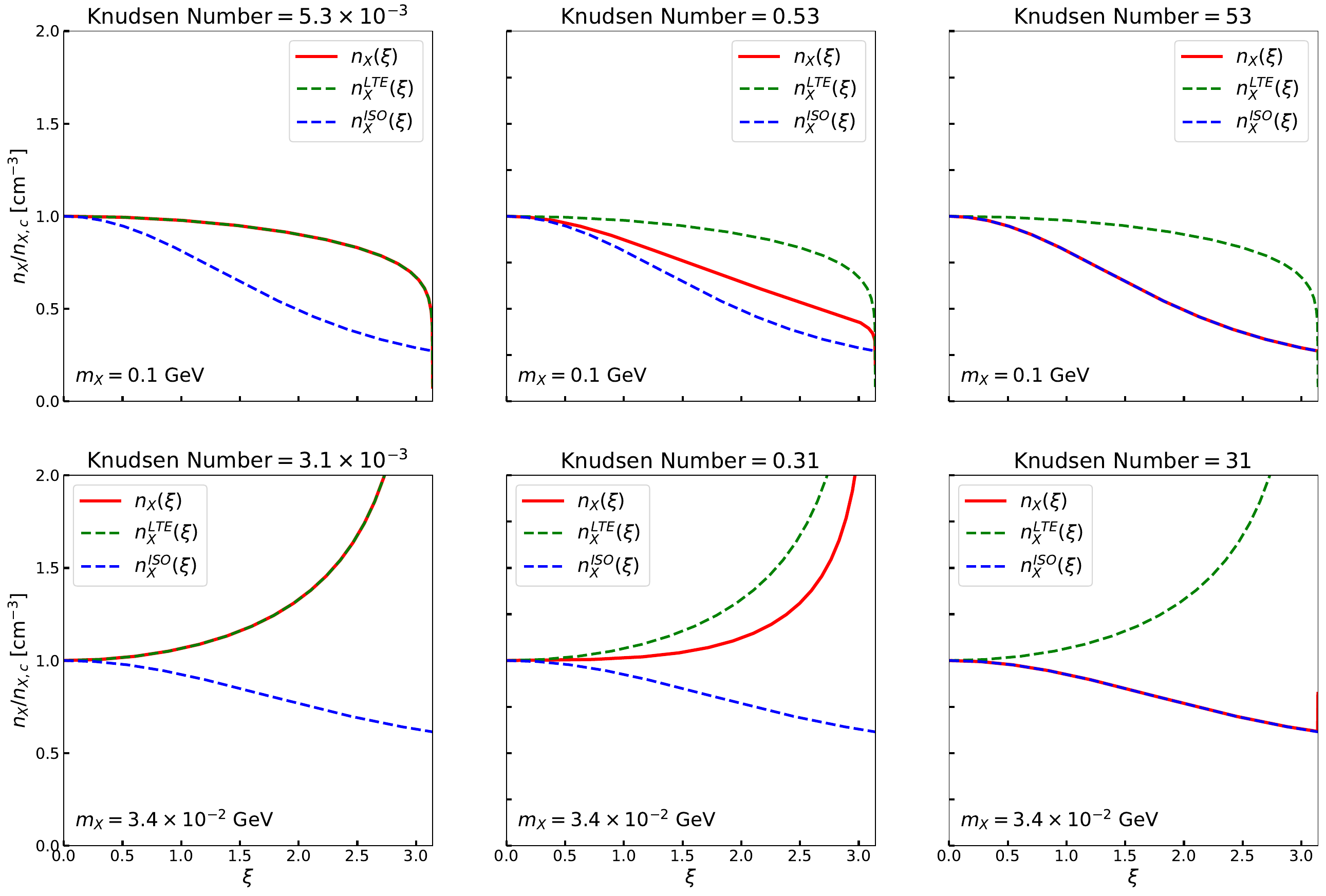}
    \caption{Dark Matter density profiles in a $1 M_J$ Jupiter for $m_X = 0.1\unit{GeV}$ (top panels) and $m_X = 3.4 \times 10^{-2}\unit{GeV}$ (bottom panels) particles in three distinct regimes: local thermal equilibrium (left panels), Knudsen transition (central panels), and isothermal (right panels). The $m_X=3.4 \times 10^{-2}\unit{GeV}$ DM particle was chosen since at this mass, $p(m_X) \sim -0.6$, making it nice to visualize the floating profile, while for $m_X = 0.1\unit{GeV}$, the profile is closer to the hydrogen profile. In each panel, the red line indicates the true DM profile given by Eq.~\ref{eq:nx_r_Transition}, while the dashed green line is the LTE profile $n_X^\text{LTE}$ as per Eq.~\ref{eq:nX-xi_LTE} and the dashed blue line is the isothermal profile $n_X^\text{ISO}$ as per Eq.~\ref{eq:nX-xi_ISO}.} 
    \label{fig:DMProfiles}
\end{figure}

Now, we shall explore the limit where $\text{Kn} \ll 1$, also known as the ``Local Thermal Equilibrium'' or LTE limit. In \cite{Gould:1990_WimpConduction}, it was shown that the DM profile in this limit is given by:
\begin{equation}
n_X(r) = n_{X}(0)\left(\frac{T(r)}{T_c}\right)^{3/2}\exp\left(-\int_0^r\frac{\alpha(r')\frac{dT}{dr'} + m_\chi \frac{d\Phi}{dr'}}{T(r')} dr'\right),
\label{eq:nxLTE_general}
\end{equation}
where $\alpha(r)$ is a factor related to the thermal diffusivity of DM. As explained in~\cite{Gould:1990_WimpConduction}, $\alpha(r)$ is actually a constant, $\alpha_0$, if we consider one nuclear species and can be approximated analytically as shown in Appendix A of~\cite{Gould:1990_WimpConduction}. With this in mind, and defining $E_e \equiv \frac{1}{2} m_X \vesc^2$, we start by making progress on the integral in the exponent:
\begin{equation}
    \begin{split}
   \int_0^r\frac{\alpha(r')\frac{dT}{dr'} + m_X \frac{d\Phi}{dr'}}{T(r')} dr' & = \int_0^\xi\frac{\alpha_0 T_c\frac{d\theta}{d\xi'} - \frac{\xi_1}{2\alpha_n}m_X \vesc^2 \frac{d\theta}{d\xi'}}{T_c \theta(\xi')} d\xi', \\
   & =\left(\alpha_0 - \frac{\xi_1}{\alpha_n}\frac{E_e}{T_c}\right)\int_0^\xi\frac{1}{\theta(\xi')}\frac{d\theta}{d\xi'}d\xi', \\
   & = \left(\alpha_0 - \frac{\xi_1}{\alpha_n}\frac{E_e}{T_c}\right)\int_0^\xi \frac{d}{d\xi'}\left[\log\left(\theta(\xi')\right)\right] d\xi', \\ 
   & = \left(\alpha_0 - \frac{\xi_1}{\alpha_n}\frac{E_e}{T_c}\right)\log\left(\theta(\xi)\right),
\end{split}
\end{equation}
 We have used $\Phi(\xi)$ for polytropes, as given by Eq.~\ref{eq:PhiXi_Polytrope} and the fact that $T\sim\theta$, as appropriate for gas pressure dominated objects, such as exoplanets.\footnote{As it turns out, $T\sim \theta$ even for Brown Dwarfs~\citep{Auddy:2016}, the other class of objects we consider in this paper.}Plugging in to Eq.~\ref{eq:nxLTE_general}, we get:
\begin{equation}
\begin{split}
n_X(\xi) & = n_X(0) \theta^{3/2}(\xi) \exp\left(-\left(\alpha_0 - \frac{\xi_1}{\alpha_n}\frac{E_e}{T_c}\right) \log\left(\theta(\xi)\right)\right)\\
& = n_X(0) \theta^{3/2}(\xi) \theta^{\frac{\xi_1}{\alpha_n}\frac{E_e}{T_c} - \alpha_0}(\xi) \\
&= n_X(0) \theta^{p}(\xi),
\end{split}
\label{eq:nX-xi_LTE}
\end{equation}
where $p \equiv \frac{\xi_1}{\alpha_n}\frac{E_e}{T_c} + \frac{3}{2} - \alpha_0$. Thus, we see that, in the LTE limit, the DM profile behaves as $n_X \sim n^p$ where $n(\xi) = n_c \theta(\xi)$ is the hydrogen number density profile and $p(m_X)$ is a mass-dependent quantity. In fact, one can show that as $m_X\to 0$, $\alpha_0 \approx 2.1$. Since $E_e\sim m_X$ and $\alpha_0>3/2$, it follows that for sufficiently small $m_X$ the exponent $p$ becomes negative. For example, assuming $n=1$, as appropriate for Jupiters, get $p\approx -0.6$, in the limit of $m_X\to 0$. This result is intriguing and somewhat counter-intuitive since as $\xi\to\xi_1$, $\theta(\xi)\to 0$ and so $n_X \to \infty$. In other words, for small enough DM mass, i.e. where $p(m_X) <0$, the profile forms a ring-like structure with higher densities toward the edge of the object. In Fig.~\ref{fig:DMProfiles} (bottom row) this phenomenon becomes clear in the plots for the density profiles for DM inside a $1 M_J$ Jupiter. A similar result was found recently for the Earth~\cite{Bramante:2022} and general celestial bodies~\cite{Leane:2022}.

Thus far, we have examined the density profile of DM particles in exoplanets in the extremes of high and low interaction. However, as is evidenced by Fig.~\ref{fig:KnudsenNumber_sigma-mx}, much of the $\sigma-m_X$ parameter space in which we are interested exists in areas where $\Kn\sim 1$. Thus, following~\cite{Garani:2017}, we approximate the true DM profile at a point in $\sigma-m_X$ parameter space via Eq.~\ref{eq:nx_r_Transition}. Note that, as expected:
$$\lim_{\Kn\to\infty} f(\Kn) = 0 \implies \lim_{\Kn\to\infty} n_X(r) = n_X^\text{ISO}(r), $$ 
and so the Isothermal profile is recovered in the Knudsen limit. Similarly:
$$\lim_{\Kn\to 0} f(\Kn) = 1 \implies \lim_{\Kn\to 0} n_X(r) = n_X^\text{LTE}(r), $$
and the local thermal equilibrium profile is recovered in the appropriate limit.

In Fig.~\ref{fig:DMProfiles}, we plot DM density profiles for a DM mass of $m_X = 0.034\unit{GeV}$ and $m_X = 0.1\unit{GeV}$ with varying values of $\sigma$ to demonstrate the three possible regimes: local thermal equilibrium, Knudsen transition, and isothermal. The range of masses were chosen to highlight the transition between a cored and floating profile in the LTE limit. In the left and right panels of this plot, we see the red line, which is the true DM profile, converge on the green and blue lines respectively, demonstrating the reduction of the full profile to the two limiting regimes of LTE and isothermal distributions. The central panels represents a scenario when $\Kn\sim 1$, and we clearly see that the true profile is a combination of the LTE and isothermal profiles. Another intriguing feature in Fig.~\ref{fig:DMProfiles} is the ``floating'' nature of the LTE profile in the bottom panel. This occurs when the exponent $p$ in Eq.~\ref{eq:nX-xi_LTE} becomes negative which, for a $1 M_J$ Jupiter, occurs for DM masses below $m_X\sim 0.04\unit{GeV}$. As mentioned in the derivation of Eq.~\ref{eq:nX-xi_LTE}, a similar result of higher DM densities towards the edge of the planet was recently shown to apply to the Earth~\cite{Bramante:2022}, though not through the use of polytropes. In~\cite{Bramante:2022}, the authors apply the same formalism for a distribution of DM that interacts frequently with the object's distribution, established in~\cite{Gould:1990_WimpConduction}, and also find a floating DM profile in the Earth.


\subsection{Dark Matter sensitivity estimates with exoplanets}\label{sec:DMSensitivity}

For self-annihilating DM models, the annihilation of captured DM particles in exoplanets could heat the object and through observations of the object's temperature, it is possible to constrain this DM model in $\sigma-m_X$ parameter space. As suggested by~\cite{Leane:2020wob}, exoplanets are excellent sub-GeV DM probes, and as such, we focus on such DM models here. In order to bypass the Lee-Weinberg limit~\cite{Lee:1977} for WIMPs ($m_X\gtrsim 1$~GeV), and still have a thermal relic for which the annihilation cross section is fixed by the relic abundance, one needs to allow for DM interacting stronger that the weak scale~\citep[e.g. see][]{Feng:2008ya}. Here we follow~\cite{Leane:2020wob} and restrict our attention to the Co-SIMP DM model of~\cite{Smirnov:2020}.

An exoplanet in thermal equilibrium will emit a luminosity $L$ at a temperature $T$ according to the Stefan-Boltzmann law:
\begin{equation}\label{eq:SBL}
    L_\text{obs} = 4\pi\epsilon\Robj^2 \sigma_\text{SB} T_\text{obs}^4,
\end{equation}
where $\sigma_\text{SB}$ is the Stefan-Boltzmann constant, and $\epsilon$ is the emisivity of the atmosphere. For simplicity, we shall assume the emissivity of the object is $\epsilon\simeq 1$. If $\epsilon < 1$, the object will become hotter by trapping outgoing heat-flow. The total flux from the exoplanet will stay constant on the basis of energy conservation, but the spectral flux density curve becomes shifted to shorter wavelengths, in view of the Wien's displacement law, and the constancy between the $\epsilon T_{obs}^4$ product. In the context of JWST searches, this is beneficial since short-wavelength bands generally have greater flux sensitivity~\cite{Leane:2020wob}.

In principle, the total luminosity comes from a combination of sources including internal heating, DM heating, and any other external heat sources. Assuming thermal equilibrium, we write the total luminosity as a sum of three components: 
$$L_\text{obs} = L_\text{int} + L_\text{DM} + L_\text{ext},$$
where $L_\text{int}\equiv 4\pi R^2\sigma_\text{SB}T_\text{int}^4$ represents the internal heating rate (from all sources except DM), $L_\text{DM}$ stands for the heating rate due to captured DM annihilations, and $L_\text{ext}$ is the rate of heating from external sources, such as the parent star. If, instead $L_\text{obs}>L_\text{int} + L_\text{DM} + L_\text{ext}$, then the object would be cooling and our sensitivity estimates become bounds. Following~\cite{Leane:2020wob}, we set $L_\text{ext} = 0$ since we are mostly interested in old, isolated exoplanets for cleaner measurements of their temperature. We note here that determining if a given exoplanet is in thermal equilibrium or cooling is difficult, particularly for isolated objects. Provided one can find a way to estimate their age, we can determine whether or not the object's temperature is above the expected value which would indicate a DM signal. What follows is valid regardless of $L_\text{ext}$, except that the sensitivity limits are instead transformed to bounds. For the internal heating, we follow~\cite{Caballero} where evolutionary models are used to obtain the temperature of old exoplanets of various masses and we interpolate to find $T_\text{int}$. This quantity ($T_{int}$) represents the exoplanet (mass dependent) effective temperature contribution from internal heat sources (excluding Dark Matter annihilations) to the object's surface temperature. One can also view $T_\text{int}$ as the minimum surface temperature the object will have, i.e. its temperature in the absence of DM heating and external heating, and we emphasize that this is not the internal temperature of the object. In appendix~\ref{sec:InternalTemp}, we outline how we estimate $T_\text{int}$ for Jupiters and Brown Dwarfs.

The final component of the exoplanet's heating comes from DM annihilation. Then, given that equilibrium between capture, annihilation, end evaporation is achieved in relatively short time-scales for the Co-SIMP models considered here~\citep{Leane:2020wob} we can express the total, constant, number of DM particles in the object as:
$$N_X = \sqrt{\frac{C}{C_A}} \frac{1}{\kappa + \frac{1}{2} E \tau_\text{eq}}.$$
To arrive at this, consider the limit of $t \gg \tau_\text{eq}/\kappa$ in Eq.~\ref{eq:Nx_t_Solved}, which gives the total number of DM particles in the object at equilibrium. Then, since $L_\text{DM} = m_X \Gamma_A = m_X C_A N_X^2$, we arrive at the following equation for the DM luminosity:
\begin{equation}
    L_\text{DM} = \frac{C m_X}{\left(\kappa + \frac{1}{2} E \tau_\text{eq}\right)^2},
\end{equation}
where we remind the reader that $\kappa \equiv \sqrt{1 + \left(\frac{1}{2} E \tau_{eq}\right)^2}$. Note that, if $E \approx 0$, then $L_\text{DM} = m_X C$, so the DM luminosity with evaporation is the DM luminosity without evaporation but suppressed by the factor $\left(\kappa + 1/2 E \tau_\text{eq}\right)^{-2}$. 

Ultimately, we are interested in solving $L_\text{obs} = L_\text{int} + L_\text{DM}$ in $\sigma - m_X$ parameter space. This equation can be recast in the following way:
\begin{equation}
     \frac{C m_X}{\left(\kappa + \frac{1}{2} E \tau_\text{eq}\right)^2} = 4\pi \Robj^2 \sigma_\text{SB} \left(T_\text{obs}^4 - T_\text{int}^4\right).
     \label{eq:BoundsEqn}
\end{equation}
We again stress that $T_\text{int}$ is not the internal temperature of the object, but rather the contribution to the object's surface temperature from any sources excluding: Dark Matter or external heating. 

At this point, we draw attention to an important and often overlooked consequence of the existence of Region II in the $\sigma-m_X$ parameter space (see Fig.~\ref{fig:AnalyticCapture_Schematic}), where the total capture rate saturates. Namely, for all $m_X$ for which DM evaporation is insignificant (compared to capture and annihilation) the constraining power of exoplanets as DM probes is completely erased if the object is hotter than a critical temperature, as we will explain below. Simply put, the lhs of Eq.~\ref{eq:BoundsEqn} becomes $\sigma$ independent, and, as such, this becomes a completely degenerate parameter. The statement above can be recast in terms of the existence of a DM and astrophysical parameters dependent critical temperature, $T_\text{crit}$, above which any astrophysical object cannot serve as a DM probe. It is straightforward to verify that the product $C m_X$ is maximized in Region II of $\sigma - m_X$ parameter space, which corresponds to a cross section so high that all DM particles crossing an object are captured, i.e. the geometric limit. Thus, let us consider Eq.~\ref{eq:BoundsEqn} in the limit that $E\to 0$ and assume we are in Region II, where the capture rate is maximized:
\begin{equation}
    C_\text{max} m_X = 4\pi \Robj^2 \sigma_\text{SB}\left(T_\text{obs;max}^4 - T_\text{int}^4\right).
    \label{eq:Ctot_constraints}
\end{equation}
 Now, since we are assuming the only heat sources are from DM annihilation and internal heat, this equation actually implies an upper bound on $T_\text{obs}$ which we label $T_\text{crit}$. For any $T_\text{obs} > T_\text{crit}$, it is impossible for the planet to be heated to the observed temperature purely by DM and internal heat sources because we have found maximal heating at $T_\text{crit}$ (by definition of $T_{crit}$). Solving for $T_\text{crit}$ by substituting Eq.~\ref{eq:CtotRII} for $C_\text{max}$ we find:
\begin{equation}
    T_\text{crit} = \left( \frac{1}{\sqrt{24\pi}} \frac{\rho_X}{\sigma_\text{SB} \vbar} \left(2\vbar^2 + 3 \vesc^2\right) + T_\text{int}^4\right)^{\frac{1}{4}}.
    \label{eq:T_Crit}
\end{equation}
Note that the critical temperature implicitly depends on the location of the exoplanet since the DM density and velocity dispersion play a key role in maximizing annihilation and are themselves dependent on location. To sum up, for every object for which we know $T_\text{int}$, its escape velocity ($\vesc$), the DM density at its location ($\rho_X$), and the DM dispersion velocity ($\vbar$) we have identified a critical temperature ($T_\text{crit}$) according to Eq.~\ref{eq:T_Crit}. If the object is observed with an inferred $T_\text{obs}>T_\text{crit}$, then we necessarily know that DM heating cannot be, alone, responsible for this overheating when compared to the expected $T_\text{int}$. This, in turn, implies the existence of external heat sources. Unless measurements are accurate enough to determine $L_\text{ext}$, one cannot, under those circumstances, make any predictions about $L_\text{DM}$, as those two are now degenerate. In turn, this implies that the object lost all of its constraining power, whenever $T_\text{obs}>T_\text{crit}$ and there is no accurate determination of $L_\text{ext}$ possible. If instead we are dealing with an isolated (rogue) exoplanet, for which $L_\text{ext}\equiv 0$ by definition, and for which one measures $T_\text{obs}>T_\text{crit}$, then this implies the existence of other unaccounted for internal heating mechanisms, {\it{excluding}} DM. Regardless, this would imply a complete loss of DM constraining power for this object.  

With the recent launch of the James Webb Space Telescope (JWST), there has been a lot of excitement about the possibility for exoplanet detection. An extensive discussion on exoplanets and their usefulness as DM probes is very nicely summarized in~\cite{Leane:2020wob}, along with useful discussions on the possibilities for detection with JWST. Included in this analysis is a determination of JWST's lowest possible temperature reach for objects located near the center of the galaxy. This limit is nicely summarized in Figure~S2 of~\cite{Leane:2020wob}. Specifically, they find that a likely JWST detection near the galactic center implies a minimum observable temperature of $T_\text{min} = 650\unit{K}$ given an exposure time of $10^5\unit{s}$. In other words, it is not possible for JWST to detect an exoplanet with an observed temperature below $T_\text{min} = 650\unit{K}$ given the telescopes sensitivity, and assuming a reasonable exposure time. We stress that this is a mere limitation of the instrument and not representative of any fundamental limit. In fact, one can show that, under the commonly made assumption of Poisson ``noise,'' we have $T_\text{min}\sim t_\text{exp}^{-1/8}$. This leads to the following exposure time dependence of $T_\text{min}$ at for  Jupiters and BDs near the galactic center:

\be\label{eq:Tmin}
T_\text{{min}}(t_\text{{exp}})=650\times \left(\frac{t_\text{{exp}}}{10^5\mathrm{s}}\right)^{-1/8}\mathrm{K}
\ee

Also note that Jupiters and Brown Dwarfs have similar radii at $\sim R_J$, so $T_\text{min}$ is object-independent. As we are focusing on planets found toward the galactic center, we thus adopt this value (i.e. 650 K) as the smallest observed temperature considered, thus assuming a hypothetical $10^5$~s exposure. For an actual observation, one can simply find the rescaled $T_\text{min}$ using Eq.~\ref{eq:Tmin}. Thus, when computing sensitivity limits to DM through exoplanet detection, we do no consider objects with observed temperature below $T_\text{min}$, since these are out of reach of JWST. Because of this, it does not make sense to consider objects whose critical temperature is below $T_\text{min}$ (i.e. $T_\text{crit} < T_\text{min}$) because if they are observed, their temperature $T_\text{obs}$ is necessarily above $T_\text{min}$, which means $T_\text{obs} > T_\text{crit}$, and, as discussed above, no constraints can be placed on the DM parameter space if this is the case. Thus, we can only consider those objects whose critical temperature surpasses the minimum observational temperature, i.e. $T_\text{crit} > 650\unit{K}$. For Jupiters around $1 M_J$, the internal heat gives rise to an expected temperature (in absence of any other heating sources) $T_\text{int}\sim 80\unit{K}$. Thus, one can verify by varying Jupiter mass in Eq.~\ref{eq:T_Crit} and taking galactocentric DM parameters (see paragraph below), that the constraint of $T_\text{crit} > T_\text{min}$ restricts us to exoplanets with mass $\Mobj \gtrsim 3 M_J$, as shown in Appendix~\ref{sec:InternalTemp}.

After this important discussion regarding the possible limitations of the usefulness of exoplanets (and astrophysical probes in general) whenever their surface temperature is larger than the critical temperature ($T_\text{crit}$) we identified, we go back to the scenarios where $T_\text{obs}<T_{crit}$. To solve Eq.~\ref{eq:BoundsEqn}, we need to have a grasp on further parameters like the DM dispersion velocity and DM density around the object. These parameters depend on the location of the exoplanet and will affect the sensitivity limits. For the purpose of being concrete, we shall consider Jupiter-like exoplanets of masses $3 M_J$ and $14 M_J$ near the galactic center and a Brown Dwarf of mass $70 M_J$ near the galactic center. The choice of the location (GC) is motivated by the high DM density, and therefore it offers the deepest reach (sensitivity) to DM any probe can have. The lower mass limit is chosen in consideration of the requirement that $T_\text{crit} > 650\unit{K}$. For each object we obtain the central temperature ($T_c$) using the numerical simulations of~\cite{Chabrier_2000}.\footnote{The values of $T_c$ estimated this way are consistent, to within a few percent, with those presented in Eqns.~3.13 and 3.16 of~\cite{Garani:2022}.} Focusing on exoplanets found near the galactic center means that for the DM dispersion velocity we have $\vbar \approx 50\unit{km/s}$ and for the DM density, $\rho_X \approx 10^3\unit{GeV/cm^3}$~\cite{Leane:2020wob}. The density does depend on the profile chosen, though they are all of order $10^3\unit{GeV/cm^3}$~\cite{Cohen_2013}.
\begin{figure}[bht]
    \centering
    \includegraphics[width=1\textwidth]{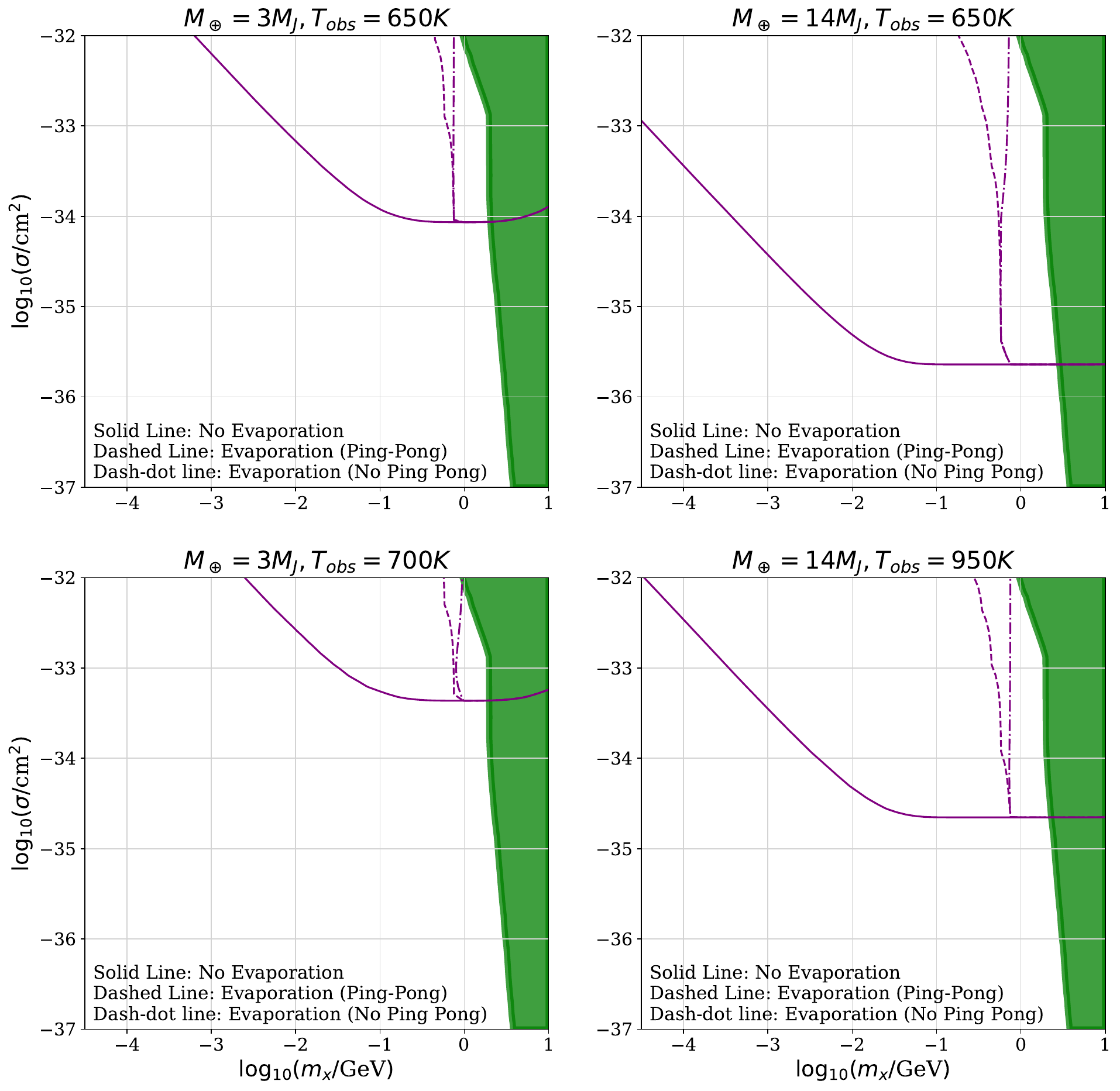}
    \caption{Sensitivity limits in $\sigma-m_X$ space for the detection of $\Mobj = 3 M_J$ (left panels) and $\Mobj = 14 M_J$ (right panels) Jupiters at temperatures $650\unit{K}$ (top panels) and $T_\text{crit}$ (bottom panels, values of $T_\text{crit}$ labeled in the title of each panel). We use a GC DM velocity and density. The dashed lines indicate the true sensitivity limits which account for evaporation, including the ping pong suppression effect of Eq.~\ref{eq:SuppressionFactor}, while the solid lines are sensitivity limits if evaporation is neglected. For consitency with previous literature, we also include sensitivity with evaporation where the suppression factor is set to $1$, as seen by the Dash-dot lines. The green region is excluded by spin-dependent (XENON1T and CDEX-1B) direct detection experiments~\cite{Aprile:2019, Aprile:2019ldm, Liu_2019}. Spin independent experiments are even more sensitive, and would have already ruled out the entire region of parameter space probed by Jupiters. and This plot demonstrates that DM evaporation severely limits the usefulness of Jupiters as sub-GeV DM probes.}
    \label{fig:SensitivityLimits}
\end{figure}
\begin{figure}[bht]
    \centering
    \includegraphics[width=1\textwidth]{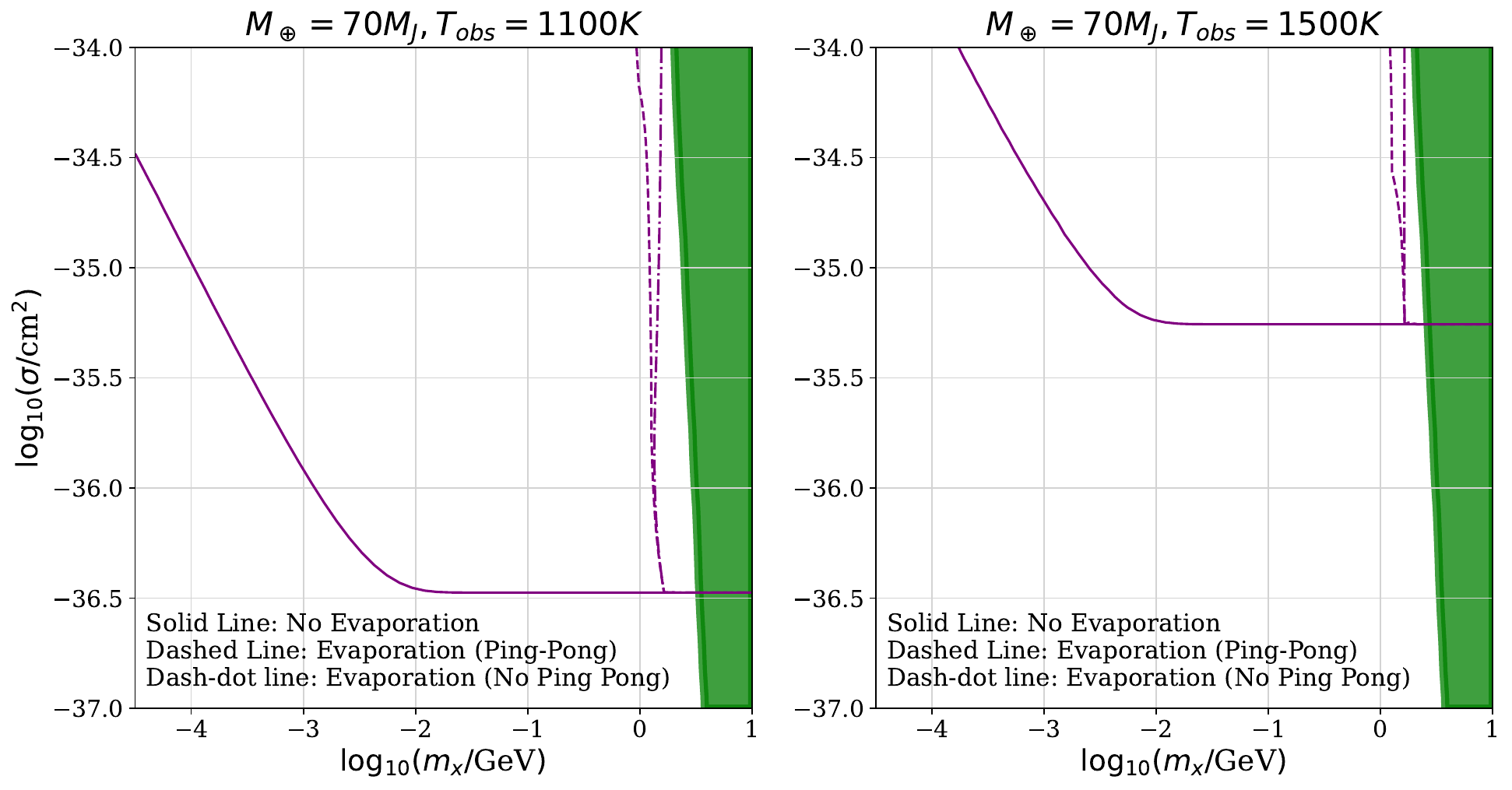}
    \caption{Sensitivity limits in $\sigma-m_X$ space for the detection of $\Mobj = 70 M_J$ Brown Dwarf at temperatures $T_{obs}=1100\unit{K}$ (left panel) and $T_{obs}\simeq T_{crit}\simeq1500\unit{K}$ (right panel). We use a GC DM velocity and density. The dashed lines indicate the true sensitivity limits which account for evaporation, including the ping-pong effect, the dash-dot lines account for evaporation without the ping-pong effect, and the solid colored lines are sensitivity limits if evaporation is neglected. The green region is excluded by spin dependent direct detection experiments. This plot demonstrates that DM evaporation severely limits the usefulness of Brown Dwarfs as sub-GeV DM probes.}
    \label{fig:SensitivityLimits_BDs}
\end{figure}

In Fig.~\ref{fig:SensitivityLimits}, we plot the sensitivity limits of two Jupiters near the galactic center, $\Mobj = 3 M_J$ (left panels) and $\Mobj = 14 M_J$ (right panels), at both the minimum temperature for JWST detection, $T_\text{min} = 650\unit{K}$ (top row), and a temperature very close to the critical temperature, $T_\text{crit}$ (bottom row). In Fig.~\ref{fig:SensitivityLimits_BDs}, we plot sensitivity limits for a $\Mobj = 70 M_J$ Brown Dwarf at temperatures $1100 K$ and $1500 K$, where the first temperature is chosen to be above the temperature due the internal heat source (i.e. $T_{int}$) while the second is just below $T_{\text{crit}}$. Near the galactic center, a $3 M_J$ Jupiter has a critical temperature $T_\text{crit} \simeq 700 \unit{K}$, a $14 M_J$ Jupiter has a critical temperature $T_\text{crit} \simeq 1000\unit{K}$, and a $70 M_J$ Brown Dwarf has a critical temperature $T_\text{crit} \simeq 1550 \unit{K}$. In both Figs.~\ref{fig:SensitivityLimits} and \ref{fig:SensitivityLimits_BDs} we consider three cases for the sensitivity limits: i. including evaporation suppressed by the $s(r)$ factor given by Eq.~\ref{eq:SuppressionFactor} (dashed lines); ii. maximum evaporation, i.e. $s(r)$ is set to one (dash-dotted lines); and iii. no evaporation considered at all (solid lines).  Since evaporation decreases the number of DM particles available for annihilation in an object, we expect that the inclusion of evaporation lowers the effect of DM heating and thus leads to weaker constraints in $\sigma-m_X$ space. This is confirmed by Figs.~\ref{fig:SensitivityLimits} and \ref{fig:SensitivityLimits_BDs}, where we see the dashed and dash-dot lines (Evaporation) diverge upwards from the solid line (No Evaporation), as evaporation begins to kick in. The fact that the constraints coincide for masses above some object-dependent cutoff is expected since evaporation is exponentially suppressed for DM masses above the ``evaporation mass.'' This quantity (the evaporation mass) is estimated for arbitrary objects in~\cite{Garani:2022}, by carefully including the effects of the tail of the Maxwell-Boltzmann distribution. This important effect was previously neglected when naive estimates for the evaporation mass were made in the literature. For $m_X$ above this mass, $E\approx 0$, and so the factor suppressing the sensitivity limits, $\left(\kappa + \frac{1}{2} E \tau_\text{eq}\right)^2$, is approximately unity, and the sensitivity with or without evaporation is identical. One important takeaway message is that DM evaporation (even when one includes its suppression) severely restricts the usefulness of Jupiters and Brown Dwarfs as sub-GeV DM probes, as can be explicitly seen from Figs.~\ref{fig:SensitivityLimits} and~\ref{fig:SensitivityLimits_BDs}. 



In what follows we discuss the expected shape of the sensitivity limits, whenever evaporation is not included (i.e. the solid lines). First, remember that those are generated as solutions to Eq.~\ref{eq:BoundsEqn}, which, in turn, depend on the region of $\sigma-m_X$ parameter space. For example, if we consider the $\Mobj = 14 M_J$ sensitivity limits for $T_\text{obs} = 650\unit{K}$ as seen in the upper right panel of Fig.~\ref{fig:SensitivityLimits}, we see that the solution below $m_X\sim 0.1\unit{GeV}$ sits at $\sigma \sim 10^{-36}\unit{cm^2}$, which one can confirm sits in Region III by computing $k$ and $\tau$ as defined in Section~\ref{sec:Capture}. Also, since we restrict our attention on the shape of the solid lines (no evaporation), the problem of finding the sensitivity limits is reduced to:
\begin{equation}
    \left(2 A \tau \vesc^2\right) m_X = 4\pi \Robj^2 \left(T_\text{obs}^4 - T_\text{int}^4 \right),
\end{equation}
where we have substituted $C$ with the analytic version in Region III given in Eq.~\ref{eq:CtotRIII}. Then, recalling that $A\sim m_X^{-1}$ and $\tau \sim \sigma$, we see that the sensitivity limits obey $\sigma \sim m_X^0$. This expectation is confirmed in all panels of Figs.~\ref{fig:SensitivityLimits} and~\ref{fig:SensitivityLimits_BDs}, where in each case one can find a flat region of the solid lines (sensitivity limits without evaporation), which corresponds to the $m_X$ range of Region~III of the $\sigma$ vs $m_X$ parameter space. We note here that the general shape of those limits is a ``bowl shaped'' curve, symmetric with respect to $m_X\sim 1$~GeV, which corresponds to the mass of the target particle inside the object, i.e. in this case protons. This symmetric shape can be traced back to the symmetry in the regions identified by us in Fig.~\ref{fig:AnalyticCapture_Schematic} with respect to the same vertical axis corresponding to $m_X\sim1$~GeV.

From Figs.~\ref{fig:SensitivityLimits} and~\ref{fig:SensitivityLimits_BDs} we also see that at either very large or very small DM mass (i.e. outside of the shallow part of the ``bowl" described above) there is some sensitivity loss, even if DM evaporation is completely negligible (see solid lines). This effect can be traced back to the difference in the capture rates in Regions III and IV, i.e. to the appearance of a factor of $k$ in Region IV when compared to Region III. As $k$ is either linear with $m_X$ (whenever $m_X<m$) or inversely proportional with $m_X$ (whenever $m_X>m$). This leads to sensitivity limits that are slanted as $\sigma\sim m_X$ or $\sigma\sim m_X^{-1}$ once we cross into Region IV or parameter space, at either end of  Region III. 

We point out that, everything else being the same, cooler objects probe deeper in to the $\sigma$ vs $m_X$ parameter space. This can be explicitly seen from contrasting for instance the top and bottom panels of Fig.~\ref{fig:SensitivityLimits} or the left and right panel of Fig.~\ref{fig:SensitivityLimits_BDs} and is a direct consequence of Eq.~\ref{eq:BoundsEqn}. As the observed temperature increases, the sensitivity limits shift upwards, until $T_{obs}\simeq T_{crit}$, when the limits depicted by the solid lines (no evaporation) in both Fig.~\ref{fig:SensitivityLimits} and~\ref{fig:SensitivityLimits_BDs} perfectly trace the boundary of Region II of parameter space, as expected. As previously explained, once $T_{obs}>T_{crit}$ the DM constraining of those objects power is lost.

\section{Summary and discussion}\label{sec:Summary}

In this work, we (re)examined the effectiveness of exoplanets and Brown Dwarfs as probes of Dark Matter, by including the effects of DM evaporation.
 Our key result is that DM evaporation, even when suppressed, almost completely erases all sensitivity at $m_X\lesssim 1\unit{GeV}$ (see Figs.~\ref{fig:SensitivityLimits} and~\ref{fig:SensitivityLimits_BDs}). While in contrast with the findings of~\cite{Leane:2020wob} (who neglect the role of DM evaporation), our results are in agreement with the findings of~\cite{Garani:2022}, who estimate the evaporation mass for arbitrary celestial bodies by including the often neglected role of the tail of the DM distribution. 
 
 Furthermore, we stress the importance of a new parameter, the critical temperature $T_\text{crit}$, that is essential to the viability a given object has a probe of annihilating dark matter via potential detection of an extra heating signal. Note that the critical temperature, as defined in Eq.~\ref{eq:T_Crit}, is not only a function of the exoplanet mass, but implicitly its location, since the DM heating signal depends on the local DM density and velocity dispersion. This upper limit on observed exoplanet temperatures arises from considering the maximum heating signal that DM can produce, corresponding to the geometric  capture limit. Simply put, exoplanets hotter than their corresponding critical temperature are not sensitive to DM since no regions in $\sigma-m_X$ space could produce a signal where $T_\text{obs} > T_\text{crit}$.\footnote{This assumes there is no external heating source, though introducing one would simply mean adding a term in the parentheses of Eq.~\ref{eq:T_Crit} and thus increasing the critical temperature.}

 The existence of the critical temperature ($T_\text{crit}$) also implies a minimum mass for a Jupiter such that it can be considered a useful DM probe. Namely, in order for an object to be detected in a reasonable exposure time, it has to have a minimum surface temperature ($T_\text{min}$). However, objects for which $T_\text{min}>T_\text{cirt}$ would not be sensitive to DM, as explained above. Conversely, one should only consider objects for which $T_\text{crit}>T_\text{min}$. As explained in Appendix~\ref{sec:InternalTemp}, this inequality can be converted into a lower bound on the mass of the DM probe. For Jupiters at the galactic center this becomes $M\gtrsim 3M_J$.

Another striking result worthy of discussion is the floating profile for light DM particles in the Local Thermal Equilibrium limit. As discussed in Section~\ref{sec:DMDistributions}, this feature arises when DM interacts frequently with the Exoplanet's hydrogen distribution and is a highly counter-intuitive result. Typically, we expect DM to develop a cored profile, with higher densities toward the center of the object. However, when the DM mass falls below a point where $p \equiv \frac{E_e}{T_c} + \frac{3}{2} - \alpha_0$ becomes negative, the DM will tend to accumulate towards the surface of the object. Similar results in recent work~\cite{Bramante:2022, Leane:2022} are exciting confirmations of this effect.

We end by summarizing our main finding. In principle, exoplanets and Brown Dwarfs could provide rich laboratories for the exploration of Dark Matter, as pointed out by~\cite{Leane:2020wob}. However, DM evaporation severely hinders their usefulness as DM probes in the poorly constrained sub-GeV regime, leaving only small swaths of the $\sigma-m_X$ parameter space that are not already ruled out by direct detection experiments. 

\appendix

\section{Analytic approximations of annihilation coefficients}\label{sec:analyticApproximations}

In this section, we use the results of Section~\ref{sec:DMDistributions} to derive analytic approximations for the annihilation and evaporation coefficients of Eq.~\ref{eq:dNx_DiffEq}. For the high-mass limit, we will compute the annihilation coefficient since the DM profiles become extremely cored in this limit, making numerical integration very difficult. We will also compute an analytic evaporation rate, though it is worth pointing out that in the high-mass limit this is not very useful since evaporation is exponentially suppressed with mass making it irrelevant above some cutoff. 

We will start by looking at integrals of the form 
$$ \int dV n_X,~\text{and}~\int dV n_X^2,$$
since these are the primary integrals we wish to approximate. We are interested in these integrals because they appear in the computation of annihilation coefficients as per Eqs.~\ref{eq:Ca_CoSIMP} and \ref{eq:Ca_2to2}, and they allow one to compute the total number of DM particles in the object. First, we remind the reader of the Taylor expansion of the Lane-Emden function $\theta(\xi)$ per Eq.~\ref{eq:thetaXi_approx}, though we only keep up to 2nd order:
$$\theta(\xi) \simeq 1 - \frac{1}{6}\xi^2.$$ 
Then, we can analytically compute the following DM distribution integral in the LTE limit:
\begin{equation}
    \begin{split}
        \int dV n_X^\text{LTE}(\xi) & = 4\pi\left(\frac{\Robj}{\xi_1}\right)^2 n_X(0)\int_0^{\xi_1} d\xi~\xi^2  \theta^p(\xi), \\
        &\simeq 4\pi\left(\frac{\Robj}{\xi_1}\right)^2 n_X(0)\int_0^{\sqrt{6}} d\xi~\xi^2  \left(1-\frac{1}{6}\xi^2\right)^p,\\
        &=  4\pi\left(\frac{\Robj}{\xi_1}\right)^2 n_X(0) \sqrt{\frac{27\pi}{2}} p^{-3/2}.
    \end{split}
    \label{eq:nxLTE_Integral}
\end{equation}
We remind the reader that $p \equiv \frac{E_e}{T_c} + \frac{3}{2} - \alpha_0$. Note how in the second line the bounds of the integral change from $0\to\xi_1$ to $0\to\sqrt{6}$, since $\sqrt{6}$ is the first zero of the approximated function and if this is not changed then, depending on $p$, we may get non-physical values for $n_X^\text{LTE}(\xi)$. In the high DM mass limit, this choice is inconsequential because the profiles become increasingly cored and so the integral converges long before $\xi = \sqrt{6}$. This will not be the case for the purely isothermal result. We can then proceed to do the same integral for the isothermal distribution:
\begin{equation}
    \begin{split}
        \int dV n_X^\text{ISO}(\xi) &= 4\pi \left(\frac{\Robj}{\xi_1}\right)^2 n_X(0) \int_0^{\xi_1} d\xi~\xi^2 \exp\left\{-\frac{m_X\vesc^2}{2 T_X}\left(1 - \theta(\xi)\right) \right\},\\
        & \simeq 4\pi \left(\frac{\Robj}{\xi_1}\right)^2 n_X(0) \int_0^{\xi_1} d\xi~\xi^2 \exp\left\{-\frac{m_X\vesc^2}{12 T_X}\xi^2 \right\}, \\
        & = 4\pi \left(\frac{\Robj}{\xi_1}\right)^2 n_X(0) \left[\frac{\sqrt{\pi}}{4 A^{3/2}} \text{Erf}\left(\sqrt{A}\xi_1\right) - \frac{\xi_1}{2 A} e^{-A \xi_1^2}\right],
    \end{split}
    \label{eq:IntegralOfnX_ISO}
\end{equation}
where:
\begin{equation}
    A\equiv \frac{m_X \vesc^2}{12 T_X}.
    \label{eq:DefineConstant_A}
\end{equation}
With these in hand, we can utilize the general form of $n_X(\xi)$ in Eq.~\ref{eq:nX-xi_FullProfile} and simply apply linearity of the integral since the Knudsen transition function, $f(\Kn)$, is radially independent:
\begin{equation}
    \int dV n_X(\xi) = f(\Kn) \int dV n_X^\text{LTE} (\xi) + \left[1 - f(\Kn)\right] \int dV n_X^\text{ISO} (\xi).
\end{equation}
Now, the more tricky approximation occurs when we look to compute the integral of $n_X^2(\xi)$. This is because of the cross term $n_X^\text{LTE} n_X^\text{ISO}.$ Let us write the expansion of $n_X^2(\xi)$:
\begin{equation}
    n_X^2(\xi) = \left[f(\Kn) n_X^\text{LTE}(\xi)\right]^2 + \left[\left(1 - f(\Kn)\right)n_X^\text{ISO}(\xi)\right]^2 + 2 f(\Kn)\left(1 - f(\Kn)\right) n_X^\text{LTE}(\xi) n_X^\text{ISO}(\xi).
\end{equation}
Thus, we are further interested in the following integrals:
$$\int dV \left[n_X^\text{LTE}(\xi)\right]^2,~\int dV \left[n_X^\text{ISO}(\xi)\right]^2,~\text{and}~ \int dV n_X^\text{LTE}(\xi) n_X^\text{ISO}(\xi).$$
The first two are straightforward. In the LTE case, we have:
\begin{equation}
\begin{split}
    \int dV \left[n_X^\text{LTE}(\xi)\right]^2 
 & \simeq 4\pi\left(\frac{\Robj}{\xi_1}\right)^2 n_X^2(0) \sqrt{\frac{27\pi}{2}} p^{-3/2}, \\
 & = \frac{n_X(0)}{\sqrt{8}} \int dV n_X^\text{LTE}(\xi).
\end{split}
\label{eq:IntegralOfnXsquared_LTE}
\end{equation}
In the isothermal limit, the integral is:
\begin{equation}
\int dV \left[n_X^\text{ISO}(\xi)\right]^2 \simeq 4\pi \left(\frac{\Robj}{\xi_1}\right)^2 n_X^2(0) \left[\frac{\sqrt{2\pi}}{16 A^{3/2}} \text{Erf}\left(\sqrt{2A}\xi_1\right) - \frac{\xi_1}{4 A} e^{- 2 A \xi_1^2}\right].
\label{eq:IntegralOfnXsquared_ISO}
\end{equation}
The cross-term integral can, remarkably, be solved analytically using the same approximation techniques as above. The result is as follows:
\begin{equation}
    \begin{split}
        \int dV n_X^\text{LTE}(\xi) n_X^\text{ISO}(\xi) & = 4\pi \left(\frac{\Robj}{\xi_1}\right)^2 n_X^2(0) \int_0^{\xi_1} d\xi~\xi^2 \theta^p(\xi) \exp\left\{-\frac{m_X\vesc^2}{2 T_X}\left(1 - \theta(\xi)\right) \right\}, \\
        & \simeq 4\pi \left(\frac{\Robj}{\xi_1}\right)^2 n_X^2(0) \int_0^{\sqrt{6}} d\xi~\xi^2 \left(1-\frac{1}{6}\xi^2\right)^p e^{-A\xi^2},\\
        & = 4\pi \left(\frac{\Robj}{\xi_1}\right)^2 n_X^2(0) \sqrt{\frac{27\pi}{2}}\frac{\Gamma(p+1)}{\Gamma\left(p+\frac{5}{2}\right)} {}_1 F_1\left(\frac{3}{2},~p+\frac{5}{2},-6 A\right),
    \end{split}
\end{equation}
where  ${}_1 F_1\left(a, b, z\right)$ is the confluent hypergeometric function of the first kind (also known as \textit{Kummer's} function), $\Gamma(z)$ is the gamma function and we remind the reader that $p \equiv \frac{E_e}{T_c} + \frac{3}{2} - \alpha_0$ and $A$ is defined in Eq.~\ref{eq:DefineConstant_A}. Note that, as with the integrals of pure LTE distributions, we cut the integral at $\xi = \sqrt{6}$ to avoid non-physical results. Again, we emphasize the fact that, in the high-mass limit, the cored nature of the profiles guarantees integral convergence before $\xi = \sqrt{6}$.


We now have all we need to compute the annihilation coefficient $C_A$ for high-mass $2\to 2$ Superheavy DM models. Recall that the annihilation coefficient is given by:
\begin{equation}
    C_A = \langle\sigma v\rangle \frac{\int dV n_X^2}{\left(\int dV n_X\right)^2}.
\end{equation}
Thus, with the above integrals completed, one can write the annihilation coefficient analytically, though it would certainly be very messy. One useful way to proceed is to understand the behavior of the coefficient in the limit of asymptotic DM mass. To do this, we shall start by looking at the asymptotic behavior of the integrals computed above. By doing so, we will actually be able to show that the asymptotic behavior is the same regardless of the Knudsen number. Starting with the LTE profile, first recall that for large $m_X$, $p\sim m_X$. Then, if we look at Eq.~\ref{eq:nxLTE_Integral} and divide out $n_X(0)$, our only mass-dependence comes from $p$, so we have:
$$\int dV \frac{n_X^\text{LTE}(\xi)}{n_X(0)} \sim p^{-3/2} \sim m_X^{-3/2}.$$
Furthermore, since the integral of $\left[n_X^\text{LTE}\right]^2$ scales like the integral of $n_X^\text{LTE}$ as per Eq.~\ref{eq:IntegralOfnXsquared_LTE}, if we divide Eq.~\ref{eq:IntegralOfnXsquared_LTE} by $n_X^2(0)$, we get the same scaling relation as above:
$$\int dV \left[\frac{n_X^\text{LTE}(\xi)}{n_X(0)}\right]^2 \sim p^{-3/2} \sim m_X^{-3/2}.$$
In the isothermal case, it is first worth recalling that as $m_X \to \infty$, $T_X \to T_c$, in other words the temperature loses its mass dependence. With this, we see that in the limit of $m_X\to\infty$, $A\sim m_X \to \infty$. Recalling the following limit:
$$\lim_{x\to\infty} \text{Erf}(x) = 1,$$
we get the following from Eq.~\ref{eq:IntegralOfnX_ISO}:
$$\lim_{m_X\to\infty} \int dV \frac{n_X^\text{ISO}(\xi)}{n_X(0)} \sim A^{-1} \left[\frac{\sqrt{\pi}}{4 A^{1/2}} - \frac{\xi_1}{2} e^{-A\xi_1^2}\right].$$
It is then clear that only the first term survives since the exponential term will decay far more quickly than the $A^{-1/2}$ term in the limit $m_X\to\infty$ (or equivalently $A\to\infty$). Thus, we obtain the following scaling relation for the integral of the isothermal profile:
$$\int dV \frac{n_X^\text{ISO}(\xi)}{n_X(0)} \sim A^{-3/2}\sim m_X^{-3/2},$$
Given the similarities of Eq.~\ref{eq:IntegralOfnX_ISO} and Eq.~\ref{eq:IntegralOfnXsquared_ISO}, we can use an identical argument to show:
$$\int dV \left[\frac{n_X^\text{ISO}(\xi)}{n_X(0)}\right]^2 \sim m_X^{-3/2}.$$
Thus, both limits have the same scaling with mass in the high mass limit. Intriguingly, the integral with the cross term also exhibits this behavior. First, we can expand \textit{Kummer's} function, ${}_1 F_1\left(a, b, z\right)$, in the limit that $z\to -\infty$. This is given by~\cite{abramowitz1965handbook}:
\begin{equation}
    \lim_{z\to-\infty}{}_1 F_1\left(a, b, z\right) = \frac{\Gamma(b)}{\Gamma(b - a)} (-z)^{-a}.
\end{equation}
Applying this to our cross term integral, we find:
\begin{equation}
        \int dV \frac{n_X^\text{LTE}(\xi) n_X^\text{ISO}(\xi)}{n_X^2(0)}  \sim \frac{\Gamma(p+1)}{\Gamma\left(p+\frac{5}{2}\right)} \frac{\Gamma\left(p+\frac{5}{2}\right)}{\Gamma(p+1)} (6 A)^{-3/2} \sim m_X^{-3/2}.
\end{equation}
With these scalings in mind, it is straightforward to show that the asymptotic behavior of $C_A$ is given by:
$$C_A \sim \langle\sigma v\rangle m_X^{3/2}.$$ 
With this, it is then possible to, for example, determine how $N_X(t \gg \tau_\text{eq})$ behaves in different regions of $\sigma-m_X$ parameter space, or determine how $n_X(0)$ behaves asymptotically with mass. These integrals are thus very useful in understanding the high-mass behavior of captured DM distributions and also allow for much easier computations in this limit as it avoids the difficulties of numerical integration for these very cored profiles. 


\section{Deriving analytic capture rates}\label{sec:AnalyticCapture}
In this appendix, we outline how we go about arriving at closed-form expressions for the DM capture rate in regions of $\sigma-m_X$ parameter space. Note that most of the difficulty in approximating the capture rate occurs in the multiscatter regime, when $\tau \gg 1$. The reason for this is that when $\tau \ll 1$, the partial capture rate for $N > 1$ is suppressed significantly, so $C_{tot} \simeq C_1$. Thus, we start with the more difficult task of approximating the multiscatter capture rate. With this in mind, we start with the general multiscatter capture rate, which we reproduce here for convenience:
\begin{equation}
    C_{N}=\frac{1}{3}\pi \Robj^{2} p_{N}(\tau) \frac{\sqrt{6} n_{X}}{\sqrt{\pi} \bar{v}}\left(\left(2 \bar{v}^{2}+3 \vesc^{2}\right)-\left(2 \bar{v}^{2}+3 v_{N}^{2}\right) \exp \left(-\frac{3\left(v_{N}^{2}-\vesc^{2}\right)}{2 \bar{v}^{2}}\right)\right),
\end{equation}
where we remind the reader of the following definitions:
$$p_N(\tau) = \frac{2}{\tau^2}\left(N+1 - \frac{\Gamma(N+2, \tau)}{N!}\right), $$
$$ v_N = \vesc\left(1 - \langle z\rangle \beta_+\right)^{-N/2},$$
$$\beta_+ = \frac{4 m m_X}{\left(m + m_X\right)^2},$$
$$k \equiv \frac{3 \vesc^2}{2 \vbar^2} \langle z\rangle \beta_+ \equiv \alpha \beta_+.$$
When considering DM-nucleon scattering cross sections that are independent of the scattering angle, we can typically make the approximation $\langle z\rangle \approx 1/2$. We also have, for all DM masses, $\beta_+ \leq 1$, which means that we can approximate $v_N$ in the following way for $\langle z\rangle \beta_+ N_{\text{max}} \ll 1$:
\begin{equation}
v_N \approx \vesc \left(1 + \frac{1}{2} \langle z\rangle \beta_+ N\right).
\label{eq:vN_approx}
\end{equation}
The derivations for the capture rates in most of $\sigma-m_X$ space relies on this approximation. However, it is very crucial to note that this approximation will not always be valid when $\alpha \ll 1$, and hence the emergence of a region without a closed form solution in Fig.~\ref{fig:AnalyticCapture_Schematic}. We will discuss this more below when we derive the capture rates in different regions. Next, we can approximate the probability function $p_N(\tau)$ in the $\tau \gg 1$ limit:
\begin{equation}
    p_N(\tau) \approx \frac{2}{\tau^2}\left(N+1\right) \Theta\left(\tau - N\right).
\end{equation}
With these approximations, we can rewrite the partial capture rate as:
\begin{equation}
    C_N \simeq \frac{2 A_X}{\tau^2} \Theta\left(\tau - N\right)\left[\left(2\vbar^2 + 3\vesc^2\right) - \left(2\vbar^2 + 3\vesc^2\right) e^{-k N} - 2 \vbar^2 k N e^{-k N}\right],
\end{equation}
where $$A_X \equiv \frac{1}{3} \pi \Robj^2 \sqrt{\frac{6}{\pi}} \frac{n_X}{\vbar}.$$
The presence of the Heaviside step function means that we need only sum $C_N$ up to $\tau$. Thus, we define the following sums for notational convenience:
\begin{equation}
    B_1 \equiv \sum_{N=1}^\tau (N+1),
\end{equation}
\begin{equation}
    B_2 \equiv \sum_{N=1}^\tau (N + 1) e^{-k N},
\end{equation}
\begin{equation}
    B_3 \equiv \sum_{N=1}^\tau N (N + 1) e^{-k N}.
\end{equation}
The total capture rate can then be expressed succinctly:
\begin{equation}
    C_{tot} \approx \frac{2 A_X}{\tau^2}\left[\left(2\vbar^2 + 3\vesc^2\right)\left(B_1 - B_2\right) - 2\vbar^2 k B_3\right].
\end{equation}
These sums can be closed and are given by:
\begin{align}
    B_1 &= \frac{1}{2}\tau(\tau+3),\label{eq:B1Ap}\\
    B_2 &= \frac{e^{-k \tau} \left(1 + e^{k\tau} \left(2 e^{k} - 1\right) - e^{k}\left(\tau + 2\right) + \tau\right)}{\left(e^k-1\right)^2},\label{eq:B2Ap}\\
    B_3 &= \frac{e^{-k \tau } \left(-e^{2 k} (\tau +2) (\tau +1)+2 e^{k (\tau +2)}+2 e^k \tau  (\tau +2)-\tau  (\tau +1)\right)}{\left(e^k-1\right)^3}\label{eq:B3Ap}.
\end{align}
We thus have completely determined the total capture rate in the multiscatter regime with two assumptions: $\langle z\rangle \beta_+ N_{\text{max}} < 1$ and $\tau \gg 1$. Note that, since the sums are being truncated at $\tau$, the first condition is the same as $\langle z\rangle \beta_+ \tau < 1$. However, in the above form, the capture rate is difficult to compute, largely because it involves exponentiation of very large numbers. Thus, we can make further assumptions, effectively dividing the $\sigma-m_X$ DM parameter space into different regions, which is what is presented in Section~\ref{sec:Capture}. 

There are two distinct cases we focus on in the multiscattering regime: $\alpha > 1$ and $\alpha < 1$. In both cases we see the emergence of Regions I and II, however there are important differences between the two and, in fact, there emerges a region in the $\alpha < 1$ regime where both approximations break down, as mentioned above when discussing the approximation of $v_N$ in Eq.~\ref{eq:vN_approx}. We will start with the $\alpha > 1$ case. The transition here surrounds the condition $k\tau \sim 1$. We define Region I as when $k\tau \ll 1$ and $\tau \gg 1$ in this case. Before proceeding further, we will show that the approximation of $v_N$ will always be true in this regime. That approximation hinged on the fact that $\langle z\rangle \beta_+ \tau \ll 1$. Note that this constraint can be rewritten as the following:
\begin{equation}
    \frac{\langle z\rangle k \tau}{\alpha} \ll 1.
    \label{eq:ConstraintForVnApprox}
\end{equation}
In this form, it is clear that when $k\tau \ll 1$ and $\alpha \geq 1$, this condition is satisfied. We also see how we might run into trouble when $\alpha < 1$. Now that we have verified the validity of the approximation we started with, we can proceed to expand around $k\tau \ll 1$ and $k\ll 1$. The reason we know $k\ll 1$ is because $\tau \gg 1$ and $k\tau \ll 1$. One can verify that by expanding they get the Region I capture rate in Eq.~\ref{eq:CtotRI}. To find the Region II capture rate, one can expand around $k \tau \gg 1$ and recover Eq.~\ref{eq:CtotRII}. A keen reader may object since we are no longer guaranteed that the approximation of $v_N$ holds. While this may be the case, we find numerically that a transition to the Region II capture rate, which is, after all, the geometric capture rate, does occur at $k\tau = 1$ (See Fig.~\ref{fig:Capture_AnalyticvsNumerical}).

Now we tackle the case when $\alpha < 1$. The key difference here is that there is no longer a transition line at $k\tau = 1$ between Regions I and II, but rather a transition region. To see why, let's re-examine Eq.~\ref{eq:ConstraintForVnApprox}. Here, we must re-define the border of Region I to be when this approximation breaks down. However, this is not the same as when $k\tau = 1$, since, atleast for $\alpha < 1/2$, when $k\tau = 1$, the condition is violated. So the region of transition occurs between the lines $k\tau = 1$ and $\frac{\langle z\rangle k \tau}{\alpha} = 1$. We do admit here that there is still a puzzle as to where exactly the transition into Region II occurs. We know that we can recover the Region II capture rate for when $k\tau \gg 1$, and this gives us an upper bound on the Region II boundary. However, numerical explorations, such as in Fig.~\ref{fig:Capture_AnalyticvsNumerical}, seem to suggest that the Region II boundary occurs somewhere below the $k\tau = 1$ line but above $\frac{\langle z\rangle k \tau}{\alpha} = 1$. We thus actually have a lower and upper bound on the Region II boundary. We do also want to point out that, even though we have yet to determine that boundary analytically, it is straightforward to use numerical techniques to estimate it.

Now we discuss the analytic approximations in the single-scatter regime, i.e.~when $\tau \ll 1$. As with the multiscatter case, there are differences between the $\alpha >1$ and $\alpha < 1$ scenarios. Specifically, Region III only emerges in the case where $\alpha > 1$. This is because when $\alpha > 1$, there are two regions to consider: $k > 1$ and $k < 1$. However, if $\alpha < 1$, then $k < 1$ always and there is only need for Region IV. To arrive at the capture rate in these regions, we make the approximation $C_{\text{tot}} \approx C_1$. This is valid because of the following expansion for $p_N(\tau)$ when $\tau \ll 1$:
\begin{equation}
    p_N(\tau) \approx \frac{2 \tau^N}{N!(N+2)},
\end{equation}
which significantly suppresses $C_N$ for any $N > 1$. Note that the approximation of $v_N$ in Eq.~\ref{eq:vN_approx} is always valid here since:
$$\langle z\rangle \beta_+ N_{\text{max}} = \langle z\rangle \beta_+ < 1.$$ We can thus write the total capture rate in the single-scatter regime as:
\begin{equation}
    C_{tot} \simeq \frac{2}{3} A_X \tau \left[\left(2\vbar^2 + 3\vesc^2\right) - \left(2\vbar^2 + 3\vesc^2\right) e^{-k} - 2 \vbar^2 k e^{-k}\right].
\end{equation}
Then, by expanding for $k \gg 1$ and $k \ll 1$, we get the capture rates Region III and IV respectively (Eqs.~\ref{eq:CtotRIII} and \ref{eq:CtotRIV}).

\section{Estimating $T_\text{int}$, i.e. the contribution of internal heating to Jupiters' and Brown Dwarfs' surface temperatures}\label{sec:InternalTemp}

In this section, we outline how we estimate the temperature of the internal heating component ($T_\text{int}$  in Eq.~\ref{eq:BoundsEqn}) of Jupiters and Brown Dwarfs. We also derive the minimum Jupiter mass for exoplanet detection in consideration of the critical temperature $T_\text{crit}$ introduced in Section~\ref{sec:DMSensitivity}. In Figure 5 of~\cite{Caballero}, evolutionary models are used to find the cooling curves of Jupiters and Brown Dwarfs. Since we are interested in exoplanets heated by DM in the late stages of their life, after they have cooled sufficiently from internal heat sources, we estimate the internal heating component of our objects of interest towards the end of their cooling curves, at an age of $\sim 1 \unit{Gyr}$. To find the temperature contribution from internal heat sources, we fit the points at the end of the cooling curve in Figure 5, where each point corresponds to a different object mass. The luminosity at these points corresponds to a time $t\sim 1\unit{Gyr}$. 
\begin{figure}[bht]
    \centering
    \includegraphics[width=1\textwidth]{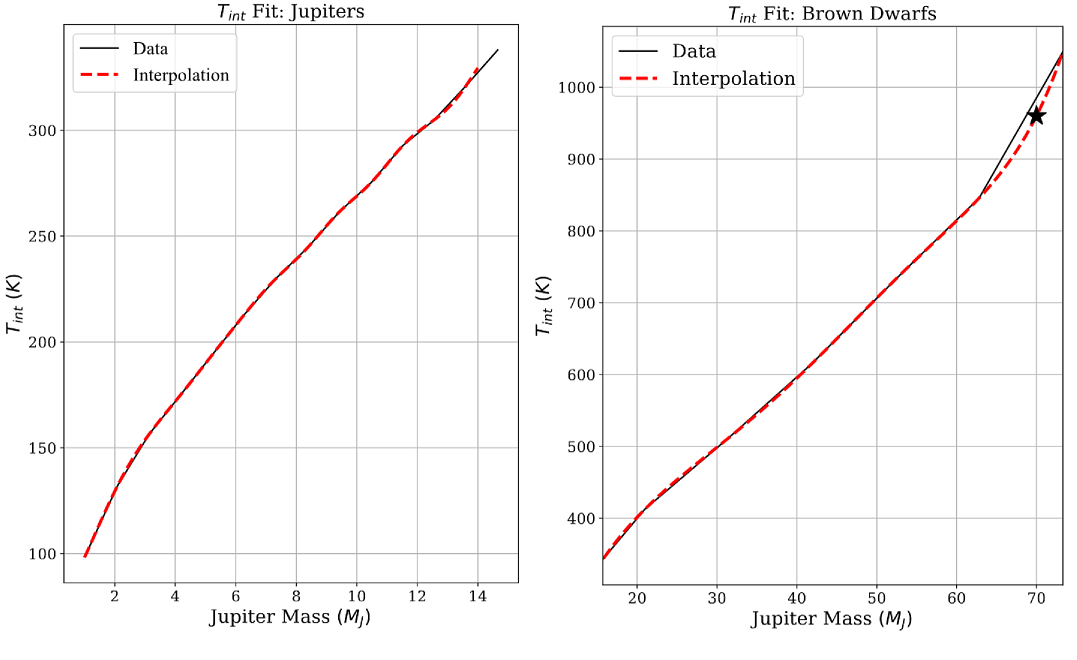}
    \caption{Interpolation of the temperature contribution to internal heating ($T_\text{int}$ in Eq.~\ref{eq:BoundsEqn}) for Jupiters and Brown Dwarfs as a function of their mass. The ``Data'' represents the temperature from the cooling curves of~\cite{Caballero} at time $\sim 1\unit{Gyr}$.}
    \label{fig:InternalTemperature}
\end{figure}
In Figure~\ref{fig:InternalTemperature}, we show the data points from Figure 5 of~\cite{Caballero}, as well as an interpolation of this data for all Jupiter and Brown Dwarf Masses.

We now go on to compute the minimum Jupiter mass to be considered as useful probes of DM for JWST searches given the critical temperature discussed in Section~\ref{sec:DMSensitivity}. As discussed there, we only consider exoplanets that can be feasibly detected by JWST in a reasonable exposure time. This implies a minimum temperature $T_\text{min} = 650\unit{K}$. We then only consider those objects with critical temperature $T_\text{crit} > T_\text{min}$, else we are not sensitive to DM signals. We thus have:
\begin{equation}
    \begin{split}
        T_\text{crit} &> T_\text{min}, \\
        \left( \frac{1}{\sqrt{24\pi}} \frac{\rho_X}{\sigma_\text{SB} \vbar} \left(2\vbar^2 + 3 \vesc^2\right) + T_\text{int}^4\right)^{\frac{1}{4}} &>T_\text{min}, \\
        \frac{1}{\sqrt{24\pi}} \frac{\rho_X}{\sigma_\text{SB} \vbar} \left(2\vbar^2 + 3 \vesc^2\right) + T_\text{int}^4 &>T_\text{min}^4, \\
        T_\text{int} &> \left(T_\text{min}^4 -  \frac{1}{\sqrt{24\pi}} \frac{\rho_X}{\sigma_\text{SB} \vbar} \left(2\vbar^2 + 3 \vesc^2\right)\right)^\frac{1}{4}.
    \end{split}
\end{equation}
Note that the object-dependence enters in both $T_\text{int}$ and $\vesc$. One can verify that the smallest mass satisfying this equation is $\Mobj \approx 3 M_J$.

\bibliographystyle{JHEP}
\bibliography{RefsDM}
\end{document}